%% file: arXiv_V2.tex
\documentclass[aps,prd,reprint, superscriptaddress, nofootinbib, hyperref={colorlinks=true,linkcolor=blue,urlcolor=blue,citecolor=blue,anchorcolor=blue}]{revtex4-1}

\usepackage{microtype}

\usepackage{graphicx}% Include figure files
\usepackage{dcolumn}% Align table columns on decimal point
\usepackage{bm}% bold math
\usepackage{lipsum}
\usepackage{amsmath,amssymb}
\usepackage[acronym,nopostdot,toc]{glossaries}
\usepackage{booktabs}
\usepackage[utf8]{inputenc}
\usepackage[T1]{fontenc}
\usepackage[normalem]{ulem} %% For striketrhough command
\usepackage{verbatim}
\usepackage[left=2cm, right=2cm, top=2cm]{geometry}
\include{abbr.tex}

\usepackage{epigraph}
\usepackage{siunitx}
\usepackage[english]{babel}
\usepackage{listings}
\usepackage{graphicx}
\usepackage[utf8]{inputenc}
\usepackage{listings}
\usepackage{matlab-prettifier}
\usepackage{amsfonts}
\usepackage{amsmath}
\usepackage{amssymb}
\usepackage{amsthm}
\usepackage{array}
\newcolumntype{P}[1]{>{\centering\arraybackslash}p{#1}}
\newcolumntype{M}[1]{>{\centering\arraybackslash}m{#1}}
\usepackage{braket}
\usepackage{xcolor}
\usepackage{enumitem}   
\usepackage{tabularx}
\usepackage{booktabs}
\usepackage{mathtools}

\usepackage{booktabs}
\usepackage{multirow}
\usepackage{orcidlink}
\usepackage{color}

\usepackage{comment}

\def\be{\begin{equation}}
\def\ee{\end{equation}}

\newcommand{\AEI}{\affiliation{Max Planck Institute for Gravitational Physics (Albert Einstein Institute) Am M\"{u}hlenberg 1, 14476 Potsdam, Germany}}
\newcommand{\Perimeter}{\affiliation{Perimeter Institute for Theoretical Physics, Ontario, N2L 2Y5, Canada}}
\newcommand{\UofG}{\affiliation{University of Guelph, Guelph, Ontario N1G 2W1, Canada}}
\newcommand{\UofGlasgow}{\affiliation{School of Physics and Astronomy, University of Glasgow, Glasgow G12 8QQ, UK}}
\newcommand{\tol}{\affiliation{Laboratoire des 2 Infinis -- Toulouse (L2IT-IN2P3), Universitè de Toulouse, CNRS, UPS, F-31062 Toulouse Cedex 9, France}}
\newcommand{\esa}{\affiliation{European Space Agency (ESA), European Space Research and Technology Centre (ESTEC), Keplerlaan 1, 2201 AZ Noordwijk, the Netherlands}}
\newcommand{\Soton}{\affiliation{School of Mathematical Sciences and STAG Research Centre, University of Southampton, Southampton, SO17 1BJ, United Kingdom}}
\newcommand{\NBI}{\affiliation{Center of Gravity, Niels Bohr Institute, Blegdamsvej 17, 2100 Copenhagen, Denmark}}

\begin{document}
\title{Systematic errors in fast relativistic waveforms for Extreme Mass Ratio Inspirals
}
\author{Hassan Khalvati
$\,$\orcidlink{0000-0001-5313-9282}}
\email{Hkhalvat@uoguelph.ca}
\Perimeter \UofG

\author{Philip Lynch $\,$\orcidlink{0000-0003-4070-7150}} 
\email{Philip.lynch@aei.mpg.de}
\AEI

\author{Ollie Burke $\,$\orcidlink{0000-0003-2393-209X}} 
\tol \UofGlasgow

\author{Lorenzo Speri $\,$\orcidlink{0000-0002-5442-7267} } \esa

\author{Maarten van de Meent \,\orcidlink{0000-0002-0242-2464}}
\NBI \AEI

\author{Zachary Nasipak $\,$\orcidlink{0000-0002-5109-9704} } 
\Soton

\begin{abstract}
Accurate modeling of \gls{EMRIs} is essential for extracting reliable information from future space-based gravitational wave observatories. Fast waveform generation frameworks 
% such as \gls{FEW} 
adopt an offline/online architecture, where expensive relativistic computations
(e.g. self-force and black hole perturbation theory) are performed offline, and waveforms are generated rapidly online via interpolation across a multidimensional parameter space. In this work, we investigate potential sources of error that result in systematic bias in these relativistic waveform models, focusing on radiation-reaction fluxes. Two key sources of systematics are identified: (i) the intrinsic inaccuracy of the flux data, for which we focus on the truncation of the multipolar mode sum, and (ii) interpolation errors from transitioning to the online stage. We quantify the impact of mode-sum truncation and analyze interpolation errors by using various grid structures and interpolation schemes. For circular orbits in Kerr spacetime with spins larger than $a \geq 0.9$, we find that $\ell_{\text{max}} \geq 30$ is required for the necessary accuracy. We also develop an efficient Chebyshev interpolation scheme, achieving the desired accuracy level with significantly fewer grid points compared to spline-based methods. For circular orbits in Kerr spacetimes, we demonstrate via Bayesian studies that interpolating the flux to a maximum global relative error that is equal to the small mass ratio is sufficient for parameter estimation purposes. For 4-year long quasi-circular EMRI signals with SNRs $= \mathcal{O}(100)$ and mass-ratios $10^{-4}-10^{-6}$, a global relative error of $10^{-6}$ yields mismatches $<10^{-3}$ and negligible parameter estimation biases.

\end{abstract}

\maketitle
% \tableofcontents

\section{Introduction}

%%% not needed for the thesis
Gravitational-wave observations provide direct access to the strong-field regime of gravity. Since the first detection by LIGO~\cite{Abbott_2016}, observations of coalescing binaries have confirmed key predictions of general relativity. The \gls{LISA}~\cite{LISA:2024hlh} will extend these tests to the milli-Hertz band, where Extreme Mass-Ratio Inspirals (\gls{EMRIs}) 
offer uniquely precise access to the strong-field region around supermassive black holes~\cite{Ryan:1997, barack2007using, glampedakis2006mapping, khalvati2025sys}.

% provide a uniquely precise probe of the spacetime around supermassive black holes~\cite{Babak:2017tow}.

%%% not needed for the thesis
EMRIs consist of a stellar-mass compact object of mass $\mu \sim 1  - 10^2 M_\odot$ inspiraling into a supermassive black hole of mass $M \sim 10^5 M_\odot - 10^7 M_\odot$ over months to years \cite{LISA:2024hlh}. Due to their extreme mass ratios of $q = \mu/M  \sim 10^{-3} - 10^{-6}$, they exhibit long durations and rich harmonic content~\cite{Drasco:2005kz, Hughes:2021exa, Babak:2017tow, Khalvati:2024tzz}. These properties make them among the most promising \gls{LISA} sources for precision tests of gravity~\cite{gair2013testing, hughes2006sort}, and probes of nuclear stellar dynamics and of possible surrounding environments, such as gas~\cite{Barausse:2007dy, Speri:2022upm, Kocsis:2011dr, Yunes:2011ws, Duque:2024mfw, Copparoni:2025jhq, Vicente:2025gsg}, dark matter~\cite{Duque_prl_dark,Speeney_2022, Mitra:2025tag}, or exotic scalar-fields~\cite{PhysRevD.101.043020,Tahura:2022ffs, Brito_scal,Dyson:2025dlj, Maselli:2020zgv,Barsanti:2022ana,Barsanti:2022vvl, Speri_2024qak, duque2024newhorizonspsiextrememassratio}. 

Extracting the science of EMRIs requires waveform models that are both highly accurate and rapid to evaluate~\cite{Strub:2025dfs, Chua:2021aah, gair2008constrained, MockLISADataChallengeTaskForce:2009wir, LISAConsortiumWaveformWorkingGroup:2023arg}. Subradian phase errors are likely needed for unbiased EMRI \gls{PE} \cite{Burke:2023lno}, while large scale data analysis pipelines require millions of waveform evaluations across a high-dimensional space. To meet these challenges, EMRIs are typically modeled within a multiscale self-force framework: the dynamics and gravitational wave emission are treated perturbatively by expanding in the small mass-ratio \cite{barack2018self,pound2022black} and by exploiting the quasi-periodic motion of the system \cite{Hinderer:2008dm,Miller:2020bft}. This naturally leads to an offline/online computational strategy. In the offline stage, expensive quantities such as waveform amplitudes \cite{Hughes:2001jr,Hughes:2005qb,Drasco:2005kz, Hughes:2021exa}, radiation-reaction fluxes \cite{Cutler:1994pb,Kennefick:1995za,Hughes:1999bq,Glampedakis:2002ya,Hughes:2005qb, Hughes:2021exa}, and self-forces \cite{Barack:2007tm, Barack:2010tm, vandeMeent:2016pee, vandeMeent:2017bcc}, are precomputed as functions of orbital parameters (e.g., semi-latus rectum, eccentricity, orbital frequencies). The online stage then rapidly constructs the waveform by interpolating this precomputed data to evolve the inspiral trajectory \cite{Warburton:2011fk, Osburn:2015duj, Warburton:2017sxk,Lynch:2021ogr,Lynch:2023gpu,Drummond:2023loz,Drummond:2023wqc,Lynch:2024ohd} and generate the gravitational wave signal~\cite{Hughes:2005qb,Drasco:2005kz, Nasipak:2023kuf}. Notably, the \gls{FEW} framework~\cite{Chua:2020stf, katz2021fast,Speri:2023jte,chapman2025fast} has combined GPU hardware acceleration with this offline/online architecture to rapidly model EMRIs in milliseconds.

An additional benefit of this perturbative approach is that higher-order information \cite{Pound:2019lzj,Warburton:2021kwk,Durkan:2022fvm,Cunningham:2024dog,Lewis:2025ydo} can be progressively incorporated to improve accuracy. So-called 0-\gls{PA} models \cite{Chua:2020stf,katz2021fast, Speri:2023jte, Nasipak:2023kuf,chapman2025fast} include only leading-order self-force effects, yielding phase errors on the order of tens of radians, whereas 1\gls{PA} models \cite{Wardell:2021fyy, Burke:2023lno} achieve subradian accuracy by adding higher-order corrections~\cite{Hinderer:2008dm, Mathews:2025nyb}. Importantly, 1\gls{PA} models build upon the 0\gls{PA} baseline. Consequently, any systematic errors in the 0\gls{PA} model propagate into the 1\gls{PA} construction \cite{Lynch:2023gpu}.

Numerical toolkits for performing the expensive offline self-force calculations are becoming more accessible and open-source~\cite{BHPToolkit, zachary_nasipak_2025_15627818, nasipak_2025_15854970, Nasipak:2025tby, Kuchler:2025hwx} and better facilitating the construction of 0\gls{PA} models in particular, such as the recent eccentric equatorial EMRI 0\gls{PA} model within the FEW framework \cite{chapman2025fast}. While this model currently lacks the pre-computed offline data to achieve full 1\gls{PA} accuracy, it remains crucial to control systematic errors in the 0\gls{PA} data so they do not overwhelm the improvements introduced by future higher-order corrections \cite{Osburn:2015duj,Lynch:2023gpu}.

One possible source of systematic errors is the interpolation of the 0PA self-force data. Due to the large number of orbital cycles expected from a typical EMRI, the 0PA terms of the equations of motion must be known to a very high degree of accuracy \cite{Burke:2021xrg}. So far, 0PA models have made use of multidimensional cubic splines trained on a dense grid of data  points to achieve the stringent accuracy requirements \cite{chua2021rapid, Hughes:2021exa, Nasipak:2023kuf,chapman2025fast,multispline}. Other EMRI related works opted instead to use the pseudo-spectral method of Chebyshev interpolation \cite{ChebyshevFunctionApproximation, boyd01} to achieve their accuracy goals while using far fewer grid points \cite{Lynch:2021ogr,Skoupy:2022adh,Lynch:2023gpu,Lynch:2024ohd}.

In this work, we quantify how the precision of precomputed 0PA self-force data---captured by radiation-reaction fluxes---and the accuracy of their interpolation, affect the orbital phase evolution and the resulting gravitational waveforms. Specifically, we investigate common, quantifiable sources of error in the model, including truncation of numerical flux sums and interpolation errors in the forcing functions, to determine the level of input accuracy required to guarantee high-fidelity waveforms for EMRI data analysis.

The Methods section (Sec.~\ref{sec:methods}) outlines our general approach for analyzing flux and phase errors, which forms the basis of all subsequent sections. 
In Sec.~\ref{sec:adiabatic_waves}, we begin by briefly reviewing the structure of the adiabatic waveforms used in this work, including how the radiation-reaction fluxes are computed from Teukolsky-based calculations. In Sec.~\ref{sec:phase and error scaling}, we present how small errors in the radiation-reaction flux can accumulate over the inspiral and turn into phase deviations. We also study how this accumulated phase error scales with various system parameters. In Sec.~\ref{sec:interp_methods}, we briefly describe the interpolation methods employed in this work, including spline interpolations and the Chebyshev-based scheme that we developed to improve efficiency and accuracy. The Sec.~\ref{sec:comparison_methods}, introduces the specific metrics and measures we use to quantify discrepancies between models in the results.

In the Results section (Sec.~\ref{sec:results_accuracy}), we present and visualize the outcomes of each analysis. The first subsection, Sec.~\ref{sec:truncation_error}, presents an analysis of errors due to truncating the multipolar sum in the flux data. We focus on a simple setup: circular equatorial orbits in Kerr spacetime. This is to keep data generation tractable, as this work requires dozens of flux datasets. We compare four different $\ell_{max}$ truncation values, summed over all $m$-modes. While our example is specific, the underlying conclusion, that truncation introduces systematic error, is general and applies beyond the circular Kerr case. 

The Sec.~\ref{sec:interpolation_error} focuses on quantifying how interpolation inaccuracies in the flux data affect the orbital phase evolution across a range of scenarios. We explore two interpolation schemes. For spline interpolation (Sec.~\ref{sec:spline_errors}), we consider different grid structures to assess their influence on both flux and phase errors. For the Chebyshev-based approach (Sec.~\ref{sec:chebyshev_errors}),
we make use of an efficient implementation which we further accelerate by pruning negligibly small coefficients while retaining a global relative accuracy (Sec.~\ref{sec:eff-cheby-interp}). In each case, we report the flux error, the resulting total accumulated phase error with respect to a reference model.

Finally, in Section~\ref{sec:interp_error_mcmc}, to assess the impact of flux interpolation errors on parameter estimation, motivated by Fig.~\ref{fig:mismatch_interp}, we select Chebyshev-based models whose mismatches
fall in the observationally relevant range, $\mathcal{M} \lesssim 0.1$ -- the regime where they may become detectable. Using these as case studies, we perform full Bayesian inference via \gls{MCMC} simulations. We quantify the extent to which interpolation-induced inaccuracies can lead to detectable biases in the recovered parameters.

\section{Methods}\label{sec:methods}

\subsection{Adiabatic waveforms for extreme mass ratio inspirals}\label{sec:adiabatic_waves}

Adiabatic waveforms model gravitational waves from EMRIs by approximating the inspiral as a slow evolution through a sequence of geodesic orbits. This relies on the radiation reaction timescale being much longer than the orbital timescale~\cite{Hinderer:2008dm, Miller:2020bft}. We follow the notation and derivations in Hughes et al.~\cite{Hughes:2021exa} based on black hole perturbation theory in the Teukolsky formalism, where the adiabatic evolution and GW strain are computed from the Weyl scalars $\psi_0$ and $\psi_4$. The stress–energy of the compact object, modeled as a point particle, acts as the source for $\psi_4$,  which at infinity is related to the GW strain via $\ddot{h} \sim \psi_4$.

For a point particle on a circular, equatorial orbit, the energy fluxes carried away from the system is computed in the Teukolsky frequency-domain framework:

\begin{align} \label{eq:flux_circular}
    &\left\langle \dot{E} \right\rangle^{\infty, H}_\text{GW} = \sum_{\ell m} \frac{1}{4\pi\omega_{m}^2} \alpha_{\ell m}^{\infty, H}\left| Z_{\ell m}^{\infty, H}\right|^2, \ 
\end{align}
where $Z^{\infty}_{\ell m}$ are Teukolsky amplitudes~\cite{Drasco:2005kz}, $\Omega_{\varphi}$ is the azimuthal orbital frequency and $\omega_{m} = m\Omega_\varphi$ are harmonic frequencies of the orbital motion. $\alpha_{\ell,m}^\infty = 1$ and $\alpha_{\ell,m}^H$ is the (mode-dependent) horizon-absorption coefficient. For quasi-circular inspirals, the angular momentum flux is obtained from $\left<\dot{L}\right>^{\infty, H}_\text{GW} = \left<\dot{E}\right>^{\infty, H}_\text{GW} \Omega_{\varphi}^{-1}$. For the explicit expression, see Ref.~\cite{Drasco:2005kz,Taracchini_2013}. These fluxes drive the inspiral by determining dissipative changes in the orbital energy and angular momentum, and hence in the orbital parameters. \\

The gravitational waveform for each fixed set of orbital parameters can be expressed as
\begin{equation}\label{eq:main_wave}
h \equiv h_+ - i h_{\times} = \frac{\mu}{d_L}\sum_{\ell m} A_{\ell m} (a,p) \ S^{\chi}_{\ell m}(\theta) \ e^{im\phi}e^{-i\omega_{m} t} , ,
\end{equation}
Here $d_L$ is the luminosity distance to the source, $(\theta, \phi)$ are polar and azimuthal view angles in the source frame, $S^{\chi}_{\ell m}$ are spin-weighted spheroidal harmonics, and the mode amplitudes are obtained from the Teukolsky amplitudes as
\begin{equation} \label{eq:amplitudes}
A_{\ell m} = -\frac{2 Z^{\infty}_{\ell m}}{\omega_{m}^2} , .
\end{equation}
 \\
% Equation \eqref{eq:main_wave} represents a snapshot for fixed orbital parameters.
In the adiabatic framework described above, $\omega_m$ and $A_{lm}$ vary slowly along the inspiral as the orbit loses energy due to radiative losses. Formally, this follows from the multiscale expansion of Einstein’s equations in the self-force formalism, which at leading order is consistent with updating the geodesic parameters in the snapshot waveform.

% \clearpage
\subsection{Phase and error accumulation for \gls{EMRIs}} \label{sec:phase and error scaling}

For \gls{EMRIs} in circular orbits, the total accumulated phase of an EMRI inspiral is given by:
\be
\Phi_{tot} = \int^T_0 \omega(t) \ dt = \int^{p_s}_{p_0} \frac{\omega(a, p)}{\dot{p}} \ dp
\ee
with $p$ is the \textit{semi-latus rectum} and represents the orbital, and $p_s$ is the \textit{separatrix}~\cite{Cutler:1994pb}. For circular orbits, they both reduce to the orbital radius $p \equiv r$, and the $p_s \equiv r_{\text{ISCO}} $. We retain the use of $p$ for consistency and generality.
For quasi-circular inspirals, the inspiral rate $\dot p$ is determined by the energy balance equation:
\be
\left(\frac{dp}{dt}\right)^{-1} = \frac{dE(a, p)}{dp} \frac{1}{\dot{E}(a, p)}
\ee
with $E(a,p)$ and $\dot{E}(a, p)$ being the total orbital energy and flux, which depend on both orbital radius $p$ and spin parameter of the central black hole $a$. We assume the standard balance law, where $\dot{E} = -\dot{E}_{GW}$. Here $\omega(a, p) \propto M^{-1}$, $E'(a,p) = dE(a,p)/dp \propto \mu M^{-1} $, and the flux $\dot{E}(a,p)$ is proportional to $ q^2$ with $q$ being the mass ratio.  Putting these together, the total phase scales as $1/q$.

If there is a deviation in the flux such that:
\be
\dot{E}_{dev}(a, p) = \dot{E}(a, p) \ (1 + \epsilon)\,
\ee
then the time evolution and phase will be perturbed by this deviation.
Such a deviation can arise from various sources, including physical effects such as environmental phenomena in EMRI systems, deviations beyond General Relativity, or actual errors and inaccuracies in the flux data. Regardless of its origin, this deviation will accumulate and affect the total phase evolution in the same manner.

The flux error is not necessarily constant and can generally depend on the orbital parameters and the spin of the central black hole. For circular orbits around a Kerr black hole, this reduces to $\epsilon \equiv \epsilon(a,p)$. Assuming $\epsilon \ll 1$  the phase error accumulates as:
\be
\Delta \Phi \simeq 
 -\int^{p_s}_{p_0} \omega(a, p)\frac{E'(a,p)}{\dot{E}(a, p)} \epsilon(a, p) \ dp
\ee
which is essentially a weighted version of the total phase $\Phi_{tot}$. 

Therefore, we can estimate the scaling of the accumulated phase error as $\Delta \Phi \propto \frac{\langle \epsilon \rangle}{q}$ where the weighted averaged flux error is defined as:
\be \label{eq:weighted error}
\langle \epsilon \rangle = \frac{\int^{p_s}_{p_0} \omega(a, p)\frac{E'(a,p)}{\dot{E}(a, p)} \epsilon(a, p) \  dp}{\Phi_{tot}} 
\ee
and for a constant flux deviation $\epsilon$, the accumulated phase deviation scales as $\Delta \Phi \sim \frac{\epsilon}{q}$. 
Near the black hole, especially closer to the ISCO, the stronger gravitational field causes orbital quantities (energy, flux, frequency) to change rapidly compared to the weak-field region farther out. This nonlinearity increases the risk of flux inaccuracies due to missing physics or numerical errors. Thus, a natural choice is to assume that $\epsilon$ scales as an inverse power law, $\epsilon(a,p) \propto g(a) p^{-n}$. While $g(a)$ depends on the spin, it acts as a constant factor during the evolution because the spin does not evolve at adiabatic order. Therefore, we focus on the dependence of $\epsilon$ on $p$, which governs how errors build up over the inspiral.

In our analysis in the following sections, we fix the observation time to the plunge, $T_{\rm plunge}$, to 4 years (as a typical observational duration~\cite{LISA:2024hlh} for the \gls{LISA} mission) when comparing different trajectories and computing phase deviations. This implies that the initial orbital radius $p_0$ will vary accordingly to ensure a consistent inspiral length to the plunge, with the plunge occurring at the $p_s \equiv r_{\text{ISCO}}$. 

Consequently, the key consideration is the width and depth of the $[p_0, p_s]$ interval within the strong-field regime and the portion of the error in the radial direction that is ingested by the EMRI.

To summarize, several factors influence the accumulated phase error. The black hole spin $a$ shifts the entire integration interval for the phase error, with higher spin moving the integration deeper into the strong-field regime, where as discussed errors are typically larger. 
% The time to plunge $T_{plunge}$ and primary mass $M$ affect the initial separation. 
The plunge time $T_{\rm plunge}$ directly affects the initial separation $p_0$ -- a shorter $T_{\rm plunge}$ corresponds to starting from a smaller $p_0$. Similarly, the primary mass $M$ affects $p_0$, with larger $M$ bringing $p_0$ closer to the ISCO due to slower inspiral ($T_r \sim \frac{M}{q}$) at a fixed mass-ratio $q$. Finally, the mass ratio $q$ introduces a dual effect. Larger $q$ leads to a faster inspiral and pushes $p_0$ further from the ISCO,  increasing the integration range and potentially accumulating more error.  However, the total number of cycles scales as $N_{\rm cycles} \sim 1/q$. This scaling dominates the accumulated phase error, as smaller $q$ results in significantly more cycles, amplifying the phase error despite the shorter integration range. Thus, the number of cycles is the primary factor in determining how the phase error scales with $q$ ($\Delta \Phi \propto \frac{\langle \epsilon \rangle}{q}$). It might seem counterintuitive that a smaller mass ratio leads to larger phase errors. We should emphasize once again that this scaling is not the accuracy of the phase itself but rather the accumulated phase error due to the known flux error $\epsilon$.

In this work, we assume $\epsilon$ to represent known error sources in the flux data and study its impact on parameter inferences. However, it is possible to infer the properties of $\epsilon$ using hierarchical inference if the phase error evolution $\Delta \Phi(t)$ can be measured accurately. This approach could provide insight into the origins of the deviations, positioning $\epsilon$ not only as a measure of flux error but also as a diagnostic tool to probe the underlying assumptions of the signals' sources.

\subsection{Interpolation methods}\label{sec:interp_methods}
Because the FEW pipeline relies on precomputed self‐force (black‐hole perturbation) data to avoid the prohibitive cost of on-the-fly evaluations, we must interpolate stored flux and amplitude values. 
% \\
An interpolation scheme for this purpose must (1) accurately reproduce the stored values at the grid points, since the data itself is assumed to be reliable within the adiabatic approximation, (2) confine any interpolation error to a local region so that inaccuracies do not propagate globally, (3) handle both weak- and strong-field regions where the function behavior may vary significantly, (4) scale efficiently to higher dimensions for generic orbits, and (5) provide smooth first and second derivatives to ensure stable evolution of the orbital phase.

\subsubsection{Spline interpolation}

To satisfy the interpolation requirements outlined earlier, we employ bicubic splines over a two-dimensional parameter space. On each rectangular cell, defined by $x_i \leq x \leq x_{i+1}$ and $y_j \leq y \leq y_{j+1}$, a bicubic spline constructs a piecewise third-degree polynomial of the form:

\be
S(x, y) = \sum_{m=0}^{3} \sum_{n=0}^{3} a_{mn} (x - x_i)^m (y - y_j)^n
\ee
with sixteen coefficients $a_{mn}$ determined from the tabulated grid values
$f_{ij}$ at $(x_i,y_j)$ and the partial derivatives $f_x, f_y, f_{xy}$ at those
points. The derivatives are computed by solving 1D cubic spline systems along rows
($y_j$ fixed) and columns ($x_i$ fixed), which enforce the continuity of the first and
second derivatives at the grid points. This structure ensures $C^2$ continuity across the entire domain, yielding a smooth surface without introducing artificial oscillations, i.e. the Runge phenomenon, associated with global high-degree fits.
The construction further requires boundary conditions for the
1D splines. The most common choice is the natural condition, where the second derivatives vanish at the endpoints. We will show later that this choice is inadequate for our purposes.

Once the spline coefficients are precomputed, each spline evaluation requires only a small local solve, making the cost \(\mathcal{O}(1)\) with respect to the global grid size, and thus is well suited for fast waveform generation. The strictly local support of the basis functions also guarantees that interpolation errors remain confined, enabling local refinements without triggering global changes. This combination of smoothness, local accuracy, and computational efficiency makes bicubic splines a natural choice for our purpose~\cite{multispline,Gezerlis_2023}. 
Spline interpolation has been used in fast waveform models~\cite{katz2021fast, chua2021rapid, Nasipak:2023kuf, Khalvati:2024tzz} and extended to higher dimensions across $(a,p,e)$ for eccentric orbits about a Kerr black hole in the new version of \gls{FEW}~\cite{chapman2025fast}.

\subsubsection{Efficient Chebyshev Interpolation} \label{sec:eff-cheby-interp}

Chebyshev interpolation \cite{ChebyshevFunctionApproximation,boyd01} is a technique used to approximate functions over an interval using a basis of Chebyshev polynomials \cite{ChebyshevPolynomials}, sampled on zeros of these polynomials called Chebyshev(-Gauss) nodes. These nodes have desirable properties such as minimizing the Runge's phenomenon, which can cause large oscillations at the edges of an interval when using high-degree polynomials for interpolation. Since this is a pseudo-spectral method, one should obtain exponential convergence in the accuracy as one increases the number of gridpoints \cite{boyd01}. This becomes particularly important when extended to functions of several variables as one can reduce the number of gridpoints required to obtain a certain target accuracy as compared with non-spectral methods. 

Our goal is to approximate a 2-dimensional function $ f(x,y)$ as a smooth continuous function of the variables $x \in [x_\text{min}, x_\text{min}]$ and $y \in [y_\text{min},  y_\text{min} ]$. One must first define a set of rescaled variables $\tilde{x},\tilde{y}  \in [-1,1]$, which we do with the simple translations
\begin{equation}
    \tilde{x} = \frac{x-(x_\text{min}+x_\text{max})/2}{(x_\text{max}-x_\text{min})/2}, \quad \tilde{y} = \frac{y-(y_\text{min}+y_\text{max})/2}{(y_\text{max}-y_\text{min})/2}.
\end{equation}

To perform Chebyshev interpolation, we sample the function at Chebyshev nodes, which are given by:
\begin{equation}
    x_i = \cos
\left(\frac{(2i + 1) \pi}{2n}\right) 
\end{equation}
for $ i = 0, 1, \ldots, n-1 $.
These nodes are the zeros of the Chebyshev polynomials of the first kind, \( T_n(x) \) which are defined recursively as:
\begin{subequations}
\begin{align}
\begin{split}
     T_0(x) &= 1, 
\end{split}\\ 
\begin{split}
     T_1(x) &= x, 
\end{split} \\
\begin{split}
     T_{n+1}(x) &= 2xT_n(x) - T_{n-1}(x). 
\end{split} 
\end{align}
\end{subequations}

One can then sample $f(\tilde{x}, \tilde{y})$ on a two dimensional grid of Chebyshev nodes with $n_x$ points in the $x$ direction and $n_y$ points in the $y$ direction. Using a discrete cosine transform (DCT) (with an appropriate renormalization), one obtains a matrix of Chebyshev coefficients $c_{i j}$ \cite{boyd01}. The resulting interpolant is given by
\begin{align} 
f(x,y) &= P_{n_x,n_y}(\tilde{x},\tilde{y}) +  \mathcal{R} \nonumber \\  &= \sum_{i=0}^{n_x-1} \sum_{j=0}^{n_y-1} c_{i j} T_i(\tilde{x}) T_j(\tilde{y}) + \mathcal{R}\label{eq:Chebyshev_Interpolant}
\end{align}
where $\mathcal{R}$ is the residual that results from truncating the Chebyshev polynomial to a finite values of $n_x$ and $n_y$. This will play an important role in our analysis in determining the size of the grid needed to obtain a desired level of accuracy.

% Accuracy paragraph
Another appealing property of Chebyshev interpolants is their in-built ability to estimate their error via the Last Coefficient Rule-of-Thumb \cite{boyd01}. First we note that $ -1 \leq T(\tilde{x}) \leq 1$ and so we can use the coefficients to estimate the overall size of the residual if one assumes that the Chebyshev polynomial is converging geometrically with the number of grid points so that $c_{i,j} \leq c_{i+1,j}$ and $ c_{i,j} \leq c_{i,j+1}$. Thus, one can put an upper bound on the residual via 
\begin{equation}
    \mathcal{R} \leq   (n_x + 1) \max_j ( | c_{n_x ,j} | ) + (n_y + 1)\max_i ( |c_{i,n_y}| )
\end{equation}
This gives an estimate of the absolute error of the interpolant, but we are often more interested in the relative error. Using the same assumptions, one can also estimate the maximum size of the  function $f(x,y)$ in the domain $x \in [x_\text{min}, x_\text{min}]$ and $y \in [y_\text{min},  y_\text{min} ]$ via 
\begin{equation} \label{eq:Chebyshev_Max_Value_Estimate}
    \max (|f|)  \sim    \max \left( |c_{i j} | \right). 
\end{equation}
As such, one can obtain a reliable estimate of the relative error $\tilde{R}$ of the interpolant via
\begin{equation}
    \tilde{R}  \leq   \frac{\left((n_x + 1) \max_j ( | c_{n_x j} | ) + (n_y + 1)\max_i ( c_{i n_y} )  \right) }{ \max \left( c_{i j} \right) }. 
\end{equation}

% Speed-up paragraph
Evaluating a two dimensional polynomial can end up being computationally prohibitive when one requires interpolants to a high degree of accuracy resulting in $n_x,n_y \sim 100$. As such, we accelerate the evaluation of our Chebyshev interpolant using the Clenshaw algorithm \cite{ClenshawAlgorithm}. In the 1D case, this reduces the $\mathcal{O}(N^2)$ operations required to evaluate the Chebyshev polynomial series of degree $N$ to $\mathcal{O}(N)$. First we evaluate the inner sum in the $\tilde{y}$ direction (i.e. over $j$) i.e.
\begin{equation} \label{eq:Chebyshev_inner_sum}
P_i(\tilde{y}) =  \sum_{j = 0}^{n_y -1} c_{ij} T_j(\tilde{y}).
\end{equation} 
This is evaluated for each value of $i$ via the Clenshaw recursion relation for Chebyshev polynomials:
\begin{equation}
    b_j = 2 \tilde{y} b_{j+1} - b_{j+1} c_{i j}
\end{equation}
with initial conditions $b_{n_y} = b_{n_y+1} = 0$. After iterating from $j = n_y - 1$ to $j = 0$ we return the result $P_i(\tilde{y}) = b_0 - b_2 \tilde{y}$.
We can then treat $P_i(\tilde{y})$ as a new set of coefficients and apply Clenshaw's algorithm once more in the $\tilde{x}$ direction (i.e. over $i$):
\begin{equation} \label{eq:Chebyshev_outer_sum}
P_{n_x,n_y}(\tilde{x},\tilde{y}) = \sum_{i = 0}^{n_x -1} P_i(\tilde{y})  T_i(\tilde{x}).
\end{equation} 
Again, we use the recurrence relation:
\begin{equation}
    b_i = 2 \tilde{x} b_{i+1} - b_{i+1} P_i(\tilde{y})
\end{equation}
with $b_{n_x} = b_{n_x+1} = 0$. After iterating from $i = n_x-1$ to $i = 0$ we return the result $P_{n_x,n_y}(\tilde{x},\tilde{y}) = b_0 - b_2 \tilde{x}$.
This means that we are only required to evaluate expressions that are linear in $\tilde{x}$ and $\tilde{y}$ at each iteration, dramatically improving computational efficiency and numerical stability.

% One can further accelerate the evaluation 

The evaluation can be made more efficient if we only require the interpolant to be accurate to a prescribed relative tolerance $\delta$. To achieve this, we apply a preprocessing step that prunes coefficients whose contributions are guaranteed to be negligible.

\begin{enumerate}
    \item We first examine the magnitudes of the coefficients $|c_{ij}|$.
Neglecting a coefficient $c_{ij}$ contributes at most an absolute error of order
$R_{ij} \lesssim (n_x+n_y)\,|c_{ij}|$.

    \item To ensure the overall truncation error remains below $\delta$, we compare this bound against the target error threshold $\delta \times \max(|c_{ij}|)$.

    \item Starting from the highest indices $(i = n_x-1, j = n_y-1)$ and moving backwards, we identify the largest index $j$ for each fixed $i$ such that
$|c_{ij}| > \frac{\delta \max(|c_{ij}|)}{n_x+n_y}$.
This defines the maximum required degree in the y-direction for that row, which we denote by $n_{y,i}(\delta)$.

    \item Finally, the effective truncation in the x-direction is set by the largest index $i$ for which $n_{y,i}(\delta) \neq 0$.
\end{enumerate}
With this scheme, the inner Chebyshev sum \(\eqref{eq:Chebyshev_inner_sum}\) only needs to be evaluated up to $n_{y,i}(\delta)$, and the outer sum \(\eqref{eq:Chebyshev_outer_sum}\) only up to the reduced x-limit. This structured pruning can dramatically reduce the number of terms required while guaranteeing a global accuracy of order $\delta$.

Finally, we note that while the above procedure was described in the two-dimensional case needed for this work, it can be extended to higher dimensional cases \footnote{Note that for a Chebyshev interpolant of dimension $d$ trained on a grid of $n^d$ points, one needs only store only $n^d$ coefficients in memory, while bicubic splines require $4^d \times (n-1)^d$ spline coefficients. This makes higher dimensional Chebyshev interpolation more memory efficient, which may prove important when one need requires higher dimensional interpolation while meeting the stringent memory requirements of current GPUs.}.

\subsection{Comparison methods}
\label{sec:comparison_methods}

Throughout this work, we employ multiple approaches to assess the accuracy of our models and identify potential systematic errors.

\subsubsection{Flux errors}
\noindent We evaluate the accuracy of our flux data by computing the fractional error relative to a reference flux data. For each model, we quantify the discrepancy as
\begin{equation}\label{eqn:error}
    \textsc{Error} = \log_{10} \left| \frac{\mathcal{F}_{\text{model}} - \mathcal{F}_{\text{ref}}}{\mathcal{F}_{\text{ref}}} \right|,
\end{equation}
where $\mathcal{F}_{\text{model}}$ is the flux values for the corresponding model and $\mathcal{F}_{\text{ref}}$ is the reference flux at each section. The results are presented as contour plots across the 2D parameter space of spin, $a$, and orbital separation, $p$, or the scaled separation parameter $u = \log{(p - p_s + 3.9)}$  which was introduced in the original FEW paper \cite{chua2021rapid}. 
\\

\subsubsection{Phase differences} 
\noindent To assess the effect of the flux error on the inspiral trajectory, we compare orbital phase evolution between each model and a reference trajectory which we specify at each section. We initialize all trajectories at an orbital separation $p_0$ such that the reference model takes 4 years to reach the plunge limit ($p_s$). We then compute the dephasing
\begin{equation}
    \Delta \Phi(t) = \Phi_{\text{model}}(t) - \Phi_{\text{ref}}(t) % \qquad \text{Dephasing}
\end{equation}
and report the final value $\Delta\Phi(t_f)$ at $t_f \equiv \min(t_{\text{plunge}, i})$, where $t_{\text{plunge}, i}$ is the plunge time of the $i$-th model in the comparison set. This approach ensures phase differences are measured at the same physical inspiral time for all cases, avoiding artifacts from comparing systems in different stages of their inspiral.
\\

\subsubsection{Mismatches} 
To assess the difference between models $(h_1,h_2)$ at the waveform level, we will use the following metrics,
\begin{align}
\mathcal{M}(h_1, h_2) &= 1 - \mathcal{O}(h_1, h_2) \,, \label{eq:mismatch} \\ %\qquad \qquad \qquad \!  \text{Mismatch} \label{eq:mismatch}\\
\mathcal{O}(h_1,h_2) &= \frac{(h_1|h_2)}{\sqrt{(h_1|h_1)(h_2|h_2)}}\,, \label{eq:overlap} \\ %\qquad \qquad  \ \text{Overlap}\label{eq:overlap}\\
(h_1|h_2) &= 4\text{Re}\int_{0}^{\infty}\text{d}f\frac{h_{1}^{\star}(f)h_{2}(f)}{S_{n}(f)}\,, \label{eq:inner_product} %\quad  \text{Inner Product}\,. \label{eq:inner_product}
\end{align}
where $\mathcal{M}$ is the mismatch, $\mathcal{O}$ is the overlap, and $(h_1|h_2)$ is the inner product.
Eq.~\eqref{eq:mismatch} quantifies orthogonality between two waveform models with respect to their amplitudes and overall phasing. The mismatch $\mathcal{M} \in [0,1]$ with $\mathcal{M} = 0$ indicating a perfect match and $\mathcal{M} = 1$ entirely orthogonal -- being as much in phase as out of phase. We define the noise-weighted inner product by Eq.\eqref{eq:inner_product} with $\mathcal{R}$ denoting the real part and $S_{n}(f)$ the Power Spectral Density \gls{PSD} of the noise process $S_{n}(f)$. The PSD governs the power of the instrumental noise as a function of frequency, suitably incorporating instrumental sensitivity into our systematic tests. 

When performing parameter estimation, we will always assume that the instrumental noise is both coloured (non-constant) and is state-of-the-art with respect to the LISA mission requirements. Specifically, we will adopt the \texttt{SciRDv1}~\cite{LISAScienceRequirementsDocument} model PSD when performing parameter estimation in the Results section \ref{sec:results_accuracy} producing Figures \ref{fig:mcmc_lmode} and (\ref{fig:MCMC_result_1eneg4} - \ref{fig:MCMC_result_1eneg6}). When performing mismatch calculations, unless specified otherwise, we will make a minor simplifying assumption that the noise process is white, resulting in a constant PSD $S_{n}(f=f_0)$ at a specific frequency. This means that the inner product \eqref{eq:inner_product} is equivalent to an integral in the time-domain via Plancherel's theorem. A result of this is that mismatches then reduce to direct comparisons between waveforms in the time-domain. Regardless of whether we conducted our analysis in the time-domain or frequency-domain (adopting non-white noise), we do not believe that our results would change significantly.

We will also define the optimal matched-filtering signal-to-noise ratio (SNR) of the waveform as the power of the waveform with respect to the variance of the noise process,
\begin{equation}
\rho = (h|h). % \qquad \text{Signal-to-noise ratio}
\end{equation}
The SNR of the signal $h$ is a measurement to bright (in power) the signal is with respect to the noise floor of the instrument determined by the \gls{PSD}.

We should note that for the purpose of pure waveform comparisons, we use mismatches with a flat \gls{PSD} = 1. However, during the Likelihood computations in the Bayesian statistics part \ref{sec:interp_error_mcmc}, we use the state-of-the-art second-generation time-delay interferometry \gls{PSD} of the LISA-detector as discussed in the following section.\\

\subsubsection{Bayesian statistics}

A gold standard technique to assess the suitability of model waveforms for parameter estimation is to use Bayesian inference. Bayes' theorem states up to a normalization constant $p(d) = \int p(d|\boldsymbol{\theta})p(\boldsymbol{\theta})\text{d}\boldsymbol{\theta}$ that 
\begin{equation}\label{eq:bayes_theorem}
    p(\boldsymbol{\theta}|d)\propto p(d|\boldsymbol{\theta})p(\boldsymbol{\theta})\,,
\end{equation}
for $p(d|\boldsymbol{\theta})$ the likelihood function, $p(\boldsymbol{\theta})$ the prior probability distribution and $p(\boldsymbol{\theta}|d)$ the sought for posterior distribution. The normalization factor $p(d) = \int p(d|\boldsymbol{\theta})p(\boldsymbol{\theta})\text{d}\boldsymbol{\theta}$ is a constant with respect to parameters and will be unnecessary for our purposes.  The goal of inference is to find parameters $\boldsymbol{\theta}$ that best reflect the observed data stream $d(t)$. One can do this by sampling auto-correlated points $\boldsymbol{\theta}$ from the posterior distribution in Eq.\eqref{eq:bayes_theorem} using Markov-Chain Monte-Carlo techniques. 

For our analysis, all details concerning the sampling algorithms and general data analysis setups are carefully described in Appendix B of Ref.~\cite{chapman2025fast}. In short, we will use the sampling algorithm \texttt{eryn}~\cite{michael_katz_2023_7705496, Katz:2024oqg} to sample from the posterior distribution in \eqref{eq:bayes_theorem}. For the analysis, we will use state-of-the-art second generation TDI variables with an accelerated \gls{LISA}-response~\cite{katz2022assessing}. The second generation \gls{PSD} of the noise process will using the \texttt{SciRDv1} model ~\cite{LISAScienceRequirementsDocument}, with the addition of the confusion background for a four year mission. We will assume that the noise is Gaussian, second-order stationary (and circulant), giving rise to a cheap-to-evaluate likelihood in the frequency domain~\cite{whittle:1957}
\begin{equation}
p(d|\boldsymbol{\theta}) \propto -\frac{1}{2}(d - h_{\text{approx}}|d - h_{\text{approx}}) \label{eq:likelihood} % \quad \text{Likelihood} \label{eq:likelihood}
\end{equation}
with $d(t) = h_{\text{true}}(t;\boldsymbol{\theta}) + n(t)$ the data stream containing the true signal $h(t;\boldsymbol{\theta})$ with parameters $\boldsymbol{\theta}$ and $n(t)$ the noise process. In Eq.\eqref{eq:likelihood}, the quantity $h_\text{approx}$ is the template model (assumed to be less accurate than the truth $h_{\text{true}}$) that is used for parameter estimation. We inject the reference waveform as the true signal into a noise-free~\footnote{This is so that we can disentangle the biases arising from systematics rather than probabilistic fluctuations to recovered parameters due to instrumental noise realizations. For farther discussion, see ~\cite{chapman_bird_2025_15630565}.} ($n(t) = 0$) data and attempt recovery using less accurate models, enabling a direct assessment of bias in parameter estimation. We will use uninformative uniform prior probability distributions as discussed in Ref.~\cite{chapman2025fast}.

\section{Results}\label{sec:results_accuracy}

\subsection{Angular mode sum truncation error} \label{sec:truncation_error}

In Sec.~\ref{sec:adiabatic_waves} we discussed that for the circular equatorial orbits, the gravitational energy flux is computed by Eq.~\ref{eq:flux_circular}, and the rate of change in angular momentum $\dot L$ is related to the energy flux $\dot E$ by $\dot L = \dot E \ \Omega_{\varphi}^{-1}$s. For brevity, we omit explicit time-averaging notation (e.g., $\left< \dot{E} \right>$), and $\dot E$ and $\dot L$ are total energy and angular momentum fluxes. \\
\begin{figure*}[t]
\centering
\includegraphics[width= \linewidth]{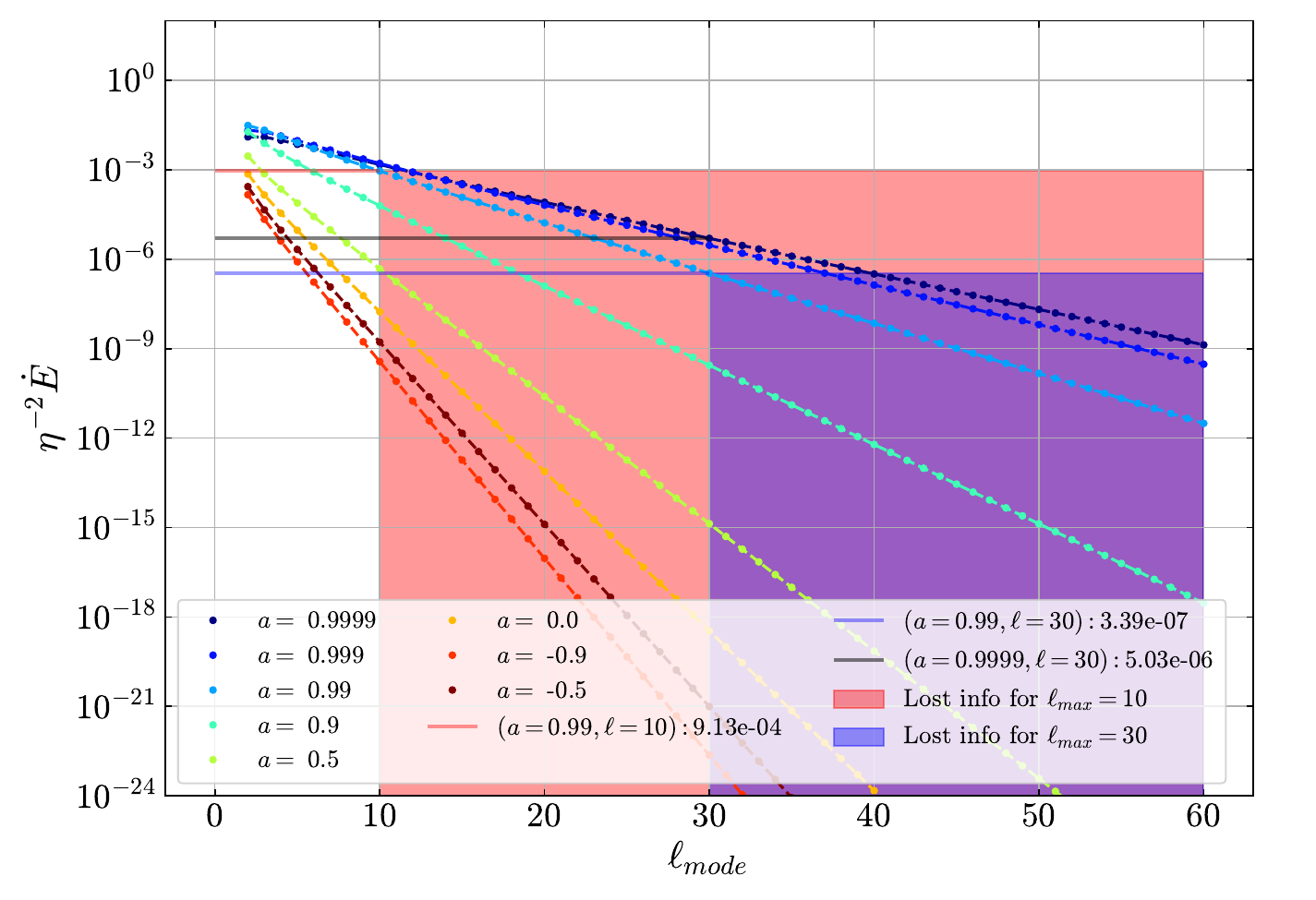}
\caption{The figure shows the contribution of each $\ell$-mode (summed over all corresponding $m$-modes) to the energy flux for different spins at $p = p_s + 0.03$. The y-axis represents the log-scaled energy flux, while the x-axis denotes $\ell$. The two shaded regions indicate the loss of information due to truncation at $\ell_{\max} = 10$ and $\ell_{\max} = 30$ as compared to $\ell_{\max} = 60$. 
The three solid lines serve as the upper bounds for truncation error beyond the corresponding  $\ell$-mode and spin value. 
% For $l_{\max} = 30$, the truncation error remains below $10^{-7}$ for the highest spin value of $a = 0.99$. 
} \label{fig:flux_mode}
\end{figure*}
The $\ell$-mode summation begins at $\ell_{\rm min} = 2$ (quadrupole radiation) and formally extends to infinity. However, in practice, only a finite number of modes are computationally feasible. Due to the \textit{super-exponential} falloff of higher angular modes~\cite{Mino_2008, Taracchini_2013}, the series can be truncated at a finite $\ell = \ell_{max}$ where $\ell_{max}$ is chosen based on the desired flux precision. A key question is: what is the convergence criterion that strikes the best balance between accuracy and computational cost?

The answer to this question hinges on ensuring the truncation error is physically irrelevant to the detector. We must translate this into quantities the detector can measure. A conservative requirement is to keep the total accumulated phase error, $\Delta\Phi$ below $\mathcal{O}(1)$~\cite{lindblom2008model, Burke:2023lno}.
% \ls{justify this choice, use references and a statment why this is conservative} 
\begin{figure*}[htbp]
\centering
\includegraphics[width= 0.95\linewidth]{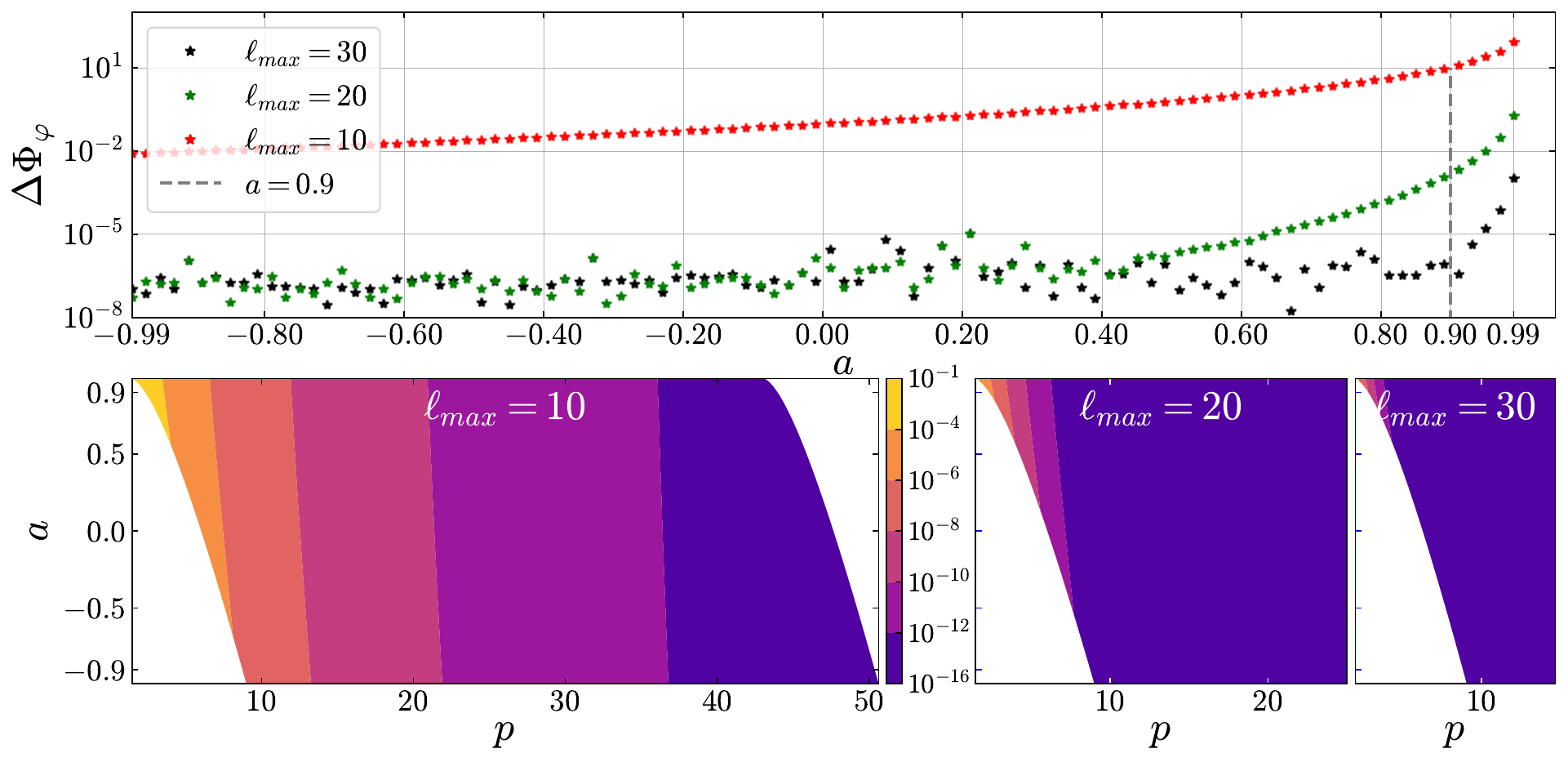}
    \caption{
    % \has{I should try p in the plot, if that looks fine, because Iam defining u here yet}
    The figure shows the phase shift and flux error contours for an EMRI system with primary mass $M = 10^6 M_{\odot}$, and secondary mass $\mu = 10 M_{\odot}$ for different primary spin values. The inspirals always start 4 years before the plunge. The top panel presents the phase shift for inspiral trajectories using fluxes truncated at $\ell_{\max} \in \{10, 20, 30\}$, assuming the correct model corresponds to $\ell_{\max} = 60$. The lower panel displays the contours of the $\log_{10}$ of flux relative error (Eq.~\ref{eqn:error}), with the reference flux having $\ell_{max}=60$ for the three choices of $\ell_{\max} = 30, 20$, and $10$.
    }
    \label{fig:flux-traj-lmodes}
\end{figure*}
\\
As derived in Section~\ref{sec:phase and error scaling}, an EMRI system accumualates $\sim \epsilon/q$ phase error during its inspiral, with $\epsilon$ being the relative error in the energy flux. As an example, for a typical EMRI $q = 10^{-5}$, we then need a choice for $\ell_{max}$ such that $\epsilon \lesssim 10^{-5}$, to get the total accumulated $\Delta\Phi \lesssim 1$.
In Fig.~\ref{fig:flux_mode}, we show the total energy flux $\dot E$ as a function of $\ell$-modes, summed over all $m$-modes. The flux is computed at $p = p_s + 0.03$ ($p_s = r_{\text{isco}}$), which is the slice in our interpolation domain where the contributions from higher modes are most significant. Results are shown for several spin values. As the spin increases, particularly for prograde orbits ($a \to 1$), the overall flux increases and becomes more concentrated in higher $\ell$-modes. This highlights the need to push $\ell_{\text{max}}$ higher for accurate modeling at high spin.
\footnote{The obviously more efficient thing to do would be to let $\ell_{\text{max}}$ depend on $a$ and $p$ using a dynamic convergence criterion, which is common practice for gathering flux for more complicated systems~(e.g., \cite{Hughes:2001jr, Skoupy:2022adh, Nasipak:2023kuf, chapman2025fast}). In this work we use a static global max as an easy-to-explain, easy-to-understand proxy of the truncation errors that occur in more sophisticated mode truncation schemes.}
%\ob{This is a good section. I think you need to say what your truthful model is (I assume $l_{\rm max} = 60$?)}

The shaded regions illustrate the flux information lost when truncating the sum at $\ell_{\text{max}} = 10$ (red) or $\ell_{\text{max}} = 30$ (purple). The solid horizontal lines show the corresponding cumulative flux values at those truncation points. Depending on the spin, truncation at $\ell = 30$ can lead to larger loss compared to lower spins, due to higher mode values. Nevertheless, even for nearly extremal spin ($a = 0.9999$), the truncation error at $\ell = 30$ remains below $\sim 10^{-6}$. We choose this near-plunge configuration as our benchmark case since it represents the most demanding region; ensuring precise and accurate flux values here guarantees sufficient accuracy at larger separations, where high-$\ell$ contributions are less significant~\cite{burke2020constraining}. We emphasize that these higher-spin cases are shown only as single-point flux diagnostics; our trajectory and interpolation models are restricted to the domain $|a|\le 0.99$.

\begin{figure}[h]
\centering
\includegraphics[width= \linewidth]{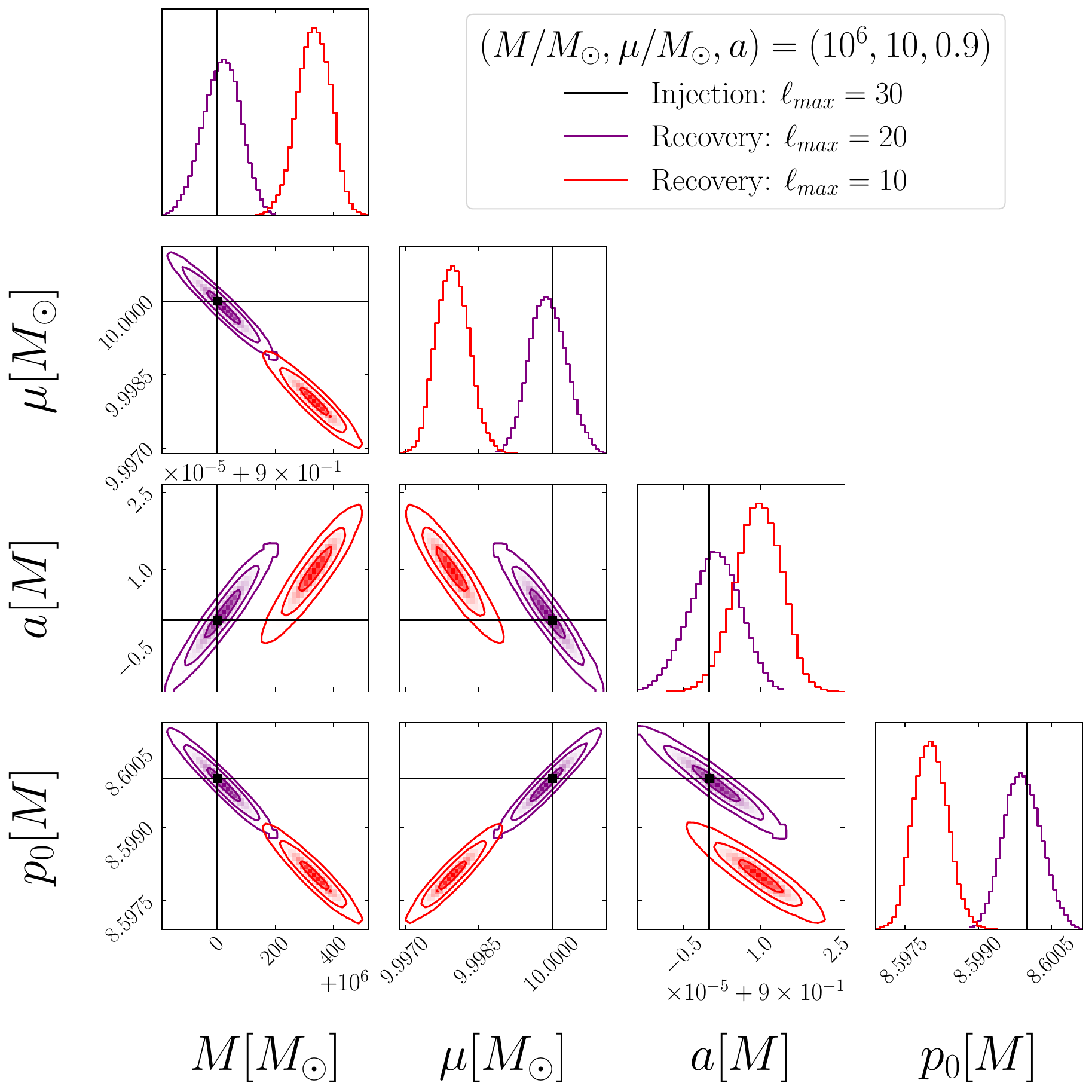}
\caption{Marginalized posterior distributions for intrinsic EMRI parameters using three different flux models with $\ell_{max} = 30$ as the true model and the $\ell_{max} = 20,10$ as approximate models. For our simulations, we consider a very bright EMRI with SNR $\sim 160$, which corresponds to a source at $d_L = 1 \text{Gpc}$.
}
\label{fig:mcmc_lmode} 
\end{figure}

To assess the overall accuracy of the flux data across the two-dimensional parameter space of $(u,a)$, we adopt the flux data computed with $\ell_{\text{max}} = 60$ as the reference model. In the lower panel of Fig.~\ref{fig:flux-traj-lmodes}, we present the relative flux error (see Eq.~\ref{eqn:error}) for truncation choices $\ell_{\text{max}} = 10$, 20, and 30. For the majority of the parameter space, the errors associated with $\ell_{\text{max}} = 20$ and 30 remain below $\sim 10^{-16}$. However, the $\ell_{\text{max}} = 10$ case shows noticeable deviation, particularly in the strong-field regime.

To examine how such flux truncation errors affect the inspiral trajectory, we simulate four inspirals using flux data with $\ell_{\text{max}} = 60$ (reference), 30, 20, and 10, for various values of spin parameter up to $a=0.99$. For all cases, we set the initial semilatus rectum $p_0$ such that the inspiral duration is approximately four years before plunge. The resulting phase shift accumulated over the inspiral is shown in the upper panel of Fig.~\ref{fig:flux-traj-lmodes}.
From the phase shift plots (upper label in Fig.~\ref{fig:flux-traj-lmodes}), we observe that the truncation error decreases systematically with increasing $\ell_{max}$. The model with $\ell_{\text{max}} = 30$ remains within a phase shift of $\lesssim 10^{-3}$ throughout the inspiral for spin parameters as large as $a=0.99$. This confirms that $\ell_{\text{max}} = 30$ is sufficiently accurate for modeling the inspiral dynamics. 

To assess potential bias in PE due to flux truncation, we perform a Bayesian inference analysis, using \gls{MCMC} sampling over the full parameter space of the system assuming the waveform with the $\ell_{max} = 30$ flux data as the accurate true model, and the two other models with $\ell_{max} = 20, 10$ as approximate models.  We choose a spin value of $a = 0.9$ as a representative moderately high-spin case. At this spin, $\ell_{\max}=20$ appears sufficient, but Fig.~\ref{fig:flux_mode} shows that in the near extremal regime ($a \sim 0.9999$) its residual error grows, making $\ell_{\max}=30$ the safer choice. We generate EMRI waveforms with parameters $(M,\mu,a,p_0) = (10^{6},10,0.9,8.6)$ resulting in a $T = 2$ year long inspiral. Incorporating the (second generation) response of the instrument, we obtain a moderately bright source SNR $\sim 160$ for a luminosity distance of $d_{L} = 1\,$Gpc. We set angular parameters to be $(\phi_S, \theta_S, \phi_K, \theta_K) = (0.5, 1.2, 0.8, 0.2)$ with initial phase $\Phi_{\phi_0} = 2$.

We present the marginalized posterior distributions for the recovered intrinsic parameters in Fig.~\ref{fig:mcmc_lmode}. As shown, the parameter inference using the approximate model waveform with $\ell_{\text{max}} = 10$ in the fluxes exhibits statistically significant parameter biases exceeding the $3\sigma$ level in multiple dimensions. This indicates that the flux error at this truncation level can lead to unreliable PE results. We do remark that the bias on the extrinsic parameters in the $\ell_{\text{max}} = 10$ case is negligible. Between the true and approximate models, the number of mode amplitudes is kept constant $|m| \leq \ell = 10$, so the main discrepancy between models is the phase evolution. The phasing is controlled by the intrinsic parameters, explaining why the main source of parameter bias is on the intrinsic parameters. It is for this reason we do not present the extrinsic parameters, but more the intrinsic ones. For our specific point in parameters pace, we found that truncating the number of $\ell$ modes to $\ell_{\rm max} = 20$ is sufficient for parameter estimation purposes.

We remind the reader that throughout this study, the waveform amplitudes are kept fixed across all cases; only the inspiral trajectories differ, as they are computed from fluxes with different $\ell_{\text{max}}$ values. This setup isolates the effect of flux-induced errors in the inspiral dynamics on parameter recovery.

\subsection{Interpolation errors} \label{sec:interpolation_error}
In Sec.\ref{sec:interp_methods} we noted that fast adiabatic FEW-style waveforms build the inspiral trajectory from the interpolation of pre-computed fluxes. We now quantify the interpolation errors and how it depends on the underlying input grid or the interpolation method.

In this section the flux error is again computed following Eq.~\ref{eqn:error} where the  $\mathcal{F}_{\text{model}}(a, u)$ is the \textbf{interpolated} flux and the $\mathcal{F}_{\text{ref}}(a, u)$ is the \textbf{exact} flux values on a relatively dense test grid with $n_u = 250$ points for $ u \in [1.36863942650, 3.82 ]$, and $n_a = 450$ for $ a \in [-0.99, +0.99]$ with uniform spacing in both dimensions. The initial point in $u$ corresponds to $p = p_s + 0.03$ which is set to be our plunge limit. The test-grid flux data, computed with $\ell_{\text{max}} = 60$, will be published alongside this work to enable verification and reuse by other researchers~\cite{khalvati_2025_17094285}. \\

Additionally, consistent with the FEW workflow, we still apply the standard scale → interpolate → rescale procedure when interpolating the flux data used in our analysis. Specifically, we remove the leading post-Newtonian term and scale it as follows~\cite{katz2021fast, chua2021rapid}:
% ~\cite{Katz_2021, few_Chua_2021}:
\begin{align}
     \dot E_\text{int} &= (\dot E_\text{grid} - \dot E^{(0)}_\text{PN} ) \ \Omega_{\varphi}^{-4} \, , \\
     \dot E^{(0)}_\text{PN} &= \frac{32}{5} \ \Omega_{\varphi}^{10/3} \, ,
  \end{align}
Here, $\dot{E}_\text{grid}$ is the raw input flux data, and $\dot{E}_\text{int}$, is the scaled flux used for interpolation. After interpolation the scaling is reversed. \\
Although the current  mid-stage scaling step effectively linearizes the data in much of the parameter space, it is not unique, and alternative scaling strategies could be explored in principle \cite{Lynch:2021ogr,Lynch:2023gpu}. However, in this work, we retain the existing scaling approach and instead keep our focus on interpolation methods and grid sparsity. Specifically, we quantify the level one can reduce the number of grid points without compromising waveform accuracy-- an important consideration when extending to higher-dimensional generic waveform models, where dense grids become computationally expensive.

\begin{figure}[t]
    \centering
    \includegraphics[width= \linewidth]{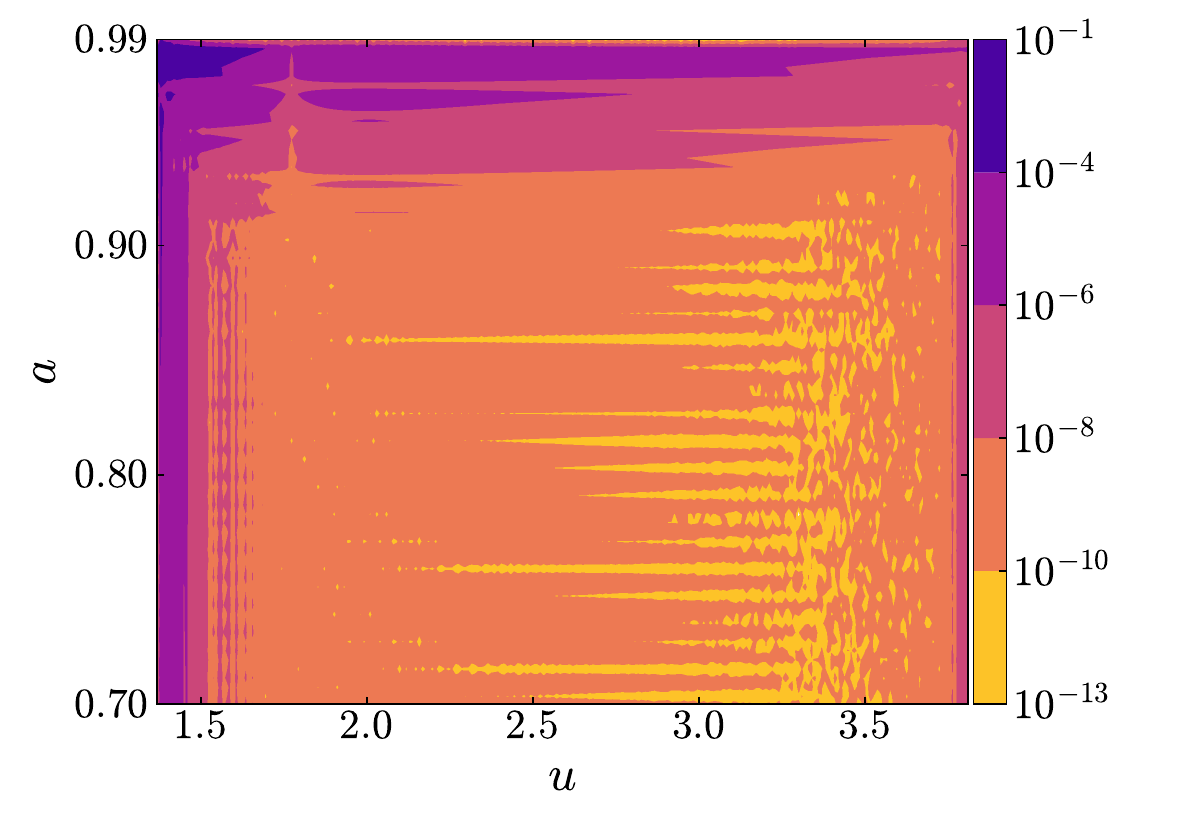}
    \caption{Interpolation error across the $(u,a)$ domain from a bicubic spline interpolant using natural boundary conditions trained on a $ 99\times 100$ uniformly spaced grid using $\Delta u = 0.025$ and $\Delta a = 0.01$ (only prograde spins of $a \in [0, 0.99]$) . The error is computed relative to the $250 \times 450$ test grid. In some regions, the error reaches levels that can accumulate over the inspiral and lead to detectable phase shifts.
}
    \label{fig:old_grid_spline}
\end{figure}

\subsubsection{Error in the Spline interpolants}\label{sec:spline_errors}
In this part we track how flux errors arising from cubic-spline interpolation accumulate into an inspiral phase shift.

We start with examining a uniform grid spacing of $\Delta u = 0.025$ and $\Delta a = 0.01$ (where here only prograde spins of $a \in [0,0.99]$ are covered), for a total of $n_u = 99$ by $n_a = 100$ points, as used in our earlier work \cite{Khalvati:2024tzz}. In Fig.~\ref{fig:old_grid_spline}, we show the interpolation error (Eq.~\ref{eqn:error}) of bicubic spline interpolants using natural boundary conditions trained on the $99 \times 100$ uniform grid against the $(250 \times 450)$ test grid. We clearly observe that the interpolation error increases significantly in the high-spin regime, confirming that a uniform spin grid introduces notable interpolation errors across the $(u,a)$ domain. Additionally, we find noticeable error at the low end of the $u$ domain, which arises from improperly set boundary conditions for the spline interpolant. These flux errors that can be large enough to be accumulated over the inspiral and lead to potential detectable phase error.  This highlights that even waveform models deemed “accurate” and fully relativistic may still carry hidden sources of error.

To mitigate this, we redesign only the spin grid, replacing the uniform spacing with a nonuniform, skewed power-law grid that clusters points more densely at high spin values (e.g., near $a = +0.99$). The $u$ grid remains unchanged, still sampled uniformly in the logarithmically scaled separation variable $u = \log(p - p_s + 3.9)$.
Because fluxes vary most rapidly and nonlinearly at high spin (Figure 1 in Ref.~\cite{Khalvati:2024tzz}), concentrating grid points in this region reduces interpolation error without altering the flux-scaling procedure. This targeted refinement improves accuracy while keeping the overall size of the grid manageable. \\

\begin{figure}[t]
\centering
\includegraphics[width= \linewidth]{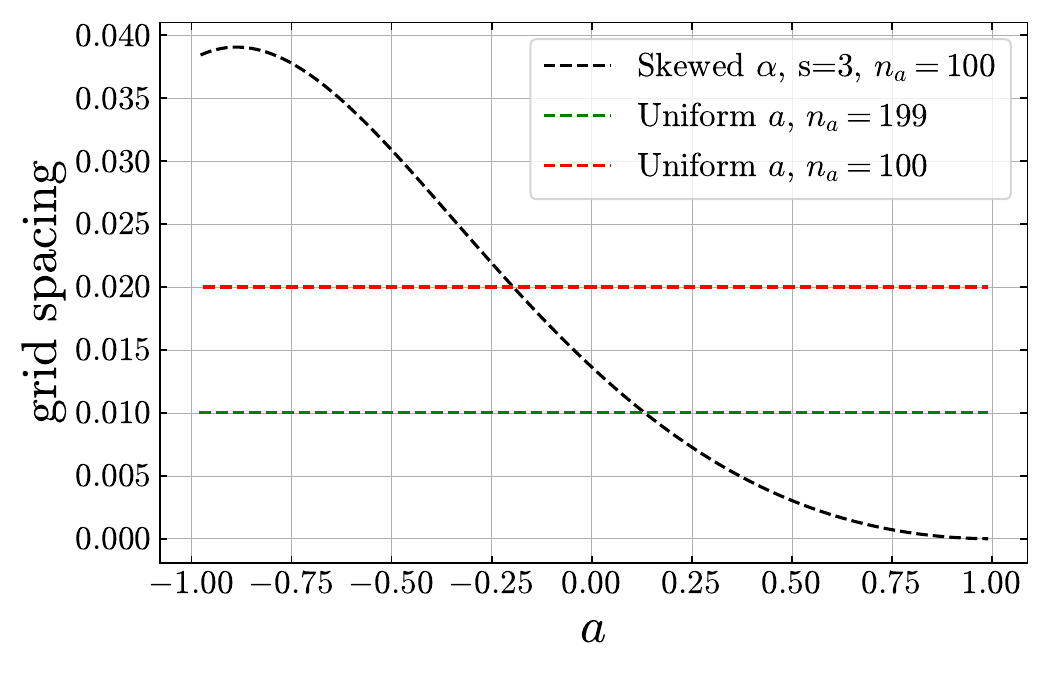}
\caption{
Grid spacing \(\Delta a\) for three different spin grid choices. The dashed black curve shows the logarithmically-scaled skewed power-law spin grid shown in Eq.~\ref{eq:alpha_grid}($s = 3$, $n_a = 100$), which concentrates resolution near high prograde spins while coarsening it near retrograde spins. This ensures that the grid spacing remains below 0.04 across the entire range. In contrast, the red and green lines represent uniform spin grids with \(N_a = 100\) and \(N_a = 199\), respectively.
} 
\label{fig:spin_grid_spacing}
\end{figure}
% \noindent\textbf{Non-Uniform Spin Grid Construction}\\
Starting with 100 uniformly spaced points in
\be \label{eq:alpha}
\alpha = \ln(a + 2.5), \qquad -0.99 \le a \le +0.99,
\ee
we skew the spacing with a power-law mapping
\be \label{eq:alpha_grid}
\alpha_{\text{grid}} = \alpha_1 + \frac{(\alpha - \alpha_1)^s}{(\alpha_2 - \alpha_1)^{s - 1}}, \qquad s = 3,
\ee
where $\alpha_1 = \ln(0.99 + 2.5)$ and $\alpha_2 = \ln(-0.99 + 2.5)$.  
This transformation retains a rectangular grid while clustering points near $a \simeq 0.99$, the region of strongest flux non-linearity. Figure~\ref{fig:spin_grid_spacing} illustrates the resulting spin-grid spacings in uniform grid cases and the skewed non-uniform case. We clearly observe the gradual increase in the grid spacing as we go towards lower spins and the retrograde cases, while always $\Delta a < 0.04$.\\
This construction maintains uniform spacing in the transformed variable $\alpha$ (i.e., fixed $\Delta\alpha$), while keeping all grid points within the physically allowed range of $-0.99 \le a \le +0.99$. Although the transformation is not unique, it serves as an illustrative example of how the choice of spin grid can affect interpolation accuracy. The present setup was selected through practical tuning: with $s = 3$ and $n_a = 100$ points, it supplies sufficient density at high spin without unduly coarsening the low-spin and retrograde regions. As such, this technique may prove to be very useful for modelling inspirals into  near-extremal black holes \cite{gralla2016inspiral,burke2020transition, burke2020constraining}.\\

\begin{figure}[t]
    \centering
    \includegraphics[width=  \linewidth]{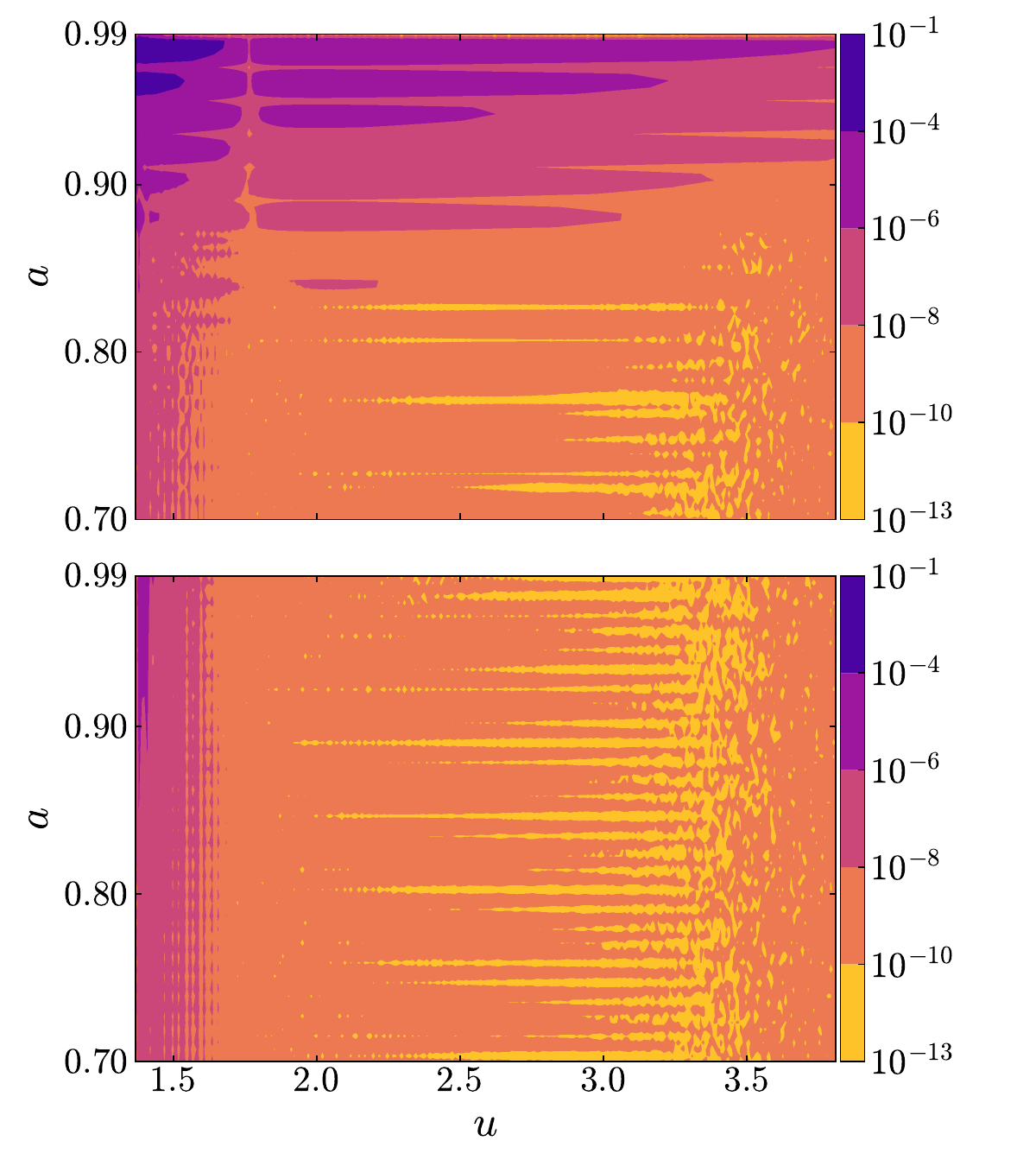}
    \caption{The relative error as measured against a $250 \times 450$ reference grid across the $(u,a)$ domain for two different interpolants, both using $n_u = 99$ points in rescaled separation and $n_a = 100$ points in spin. The top panel shows the result for a uniform grid with fixed spacing $\Delta a = 0.02$, while the bottom panel uses the non-uniform, skewed power-law grid introduced in Eq.~\ref{eq:alpha_grid}. The plotted region focuses on high spin ($a \geq 0.7$), where interpolation error is most relevant; in the lower-spin regime, the error remains negligible and follows a similar trend in both cases.
}
    \label{fig:interp_error_uniform_vs_skewed}
\end{figure}

We compute the flux data for each value of $u$ and $\alpha$ across the non-uniform spin grid defined in Eq.~\ref{eq:alpha_grid} with $n_u = 99$ and $n_\alpha= 100$. The flux is then interpolated with a bicubic spline over the transformed variables $\alpha$, (Eq.~\ref{eq:alpha}) and $u$. As a point of comparison, we also construct a second uniform grid which now also includes retrograde spins so that $a \in [-0.99,0.99] $. We use a larger spacing of $\Delta a = 0.02$ so that the grid maintains the same size of $99\times 100$ points. This is also interpolated using the same bicubic spline method.

Figure~\ref{fig:interp_error_uniform_vs_skewed} shows the resulting interpolation error across the $(u,a)$ for both the uniform (top panel) and non-uniform (bottom panel) spin grids. The relative error is computed following Eq.~\ref{eqn:error}, where the reference flux $\mathcal{F}_{\text{ref}}$ corresponds to the values on our $250 \times 450$ test grid data set.
 As seen in the top panel, the uniform spacing of $\delta a = 0.02$ is insufficient in the high-spin regime, where the flux becomes increasingly non-linear. The interpolation error rises sharply for $a \gtrsim 0.9$, indicating poor performance in this region. In contrast, the non-uniform grid in the bottom panel—while using the same number of points—significantly reduces the error at high spin by concentrating more grid points where the flux varies most rapidly.\\

\begin{figure}[t]
\centering
\includegraphics[width= \linewidth]{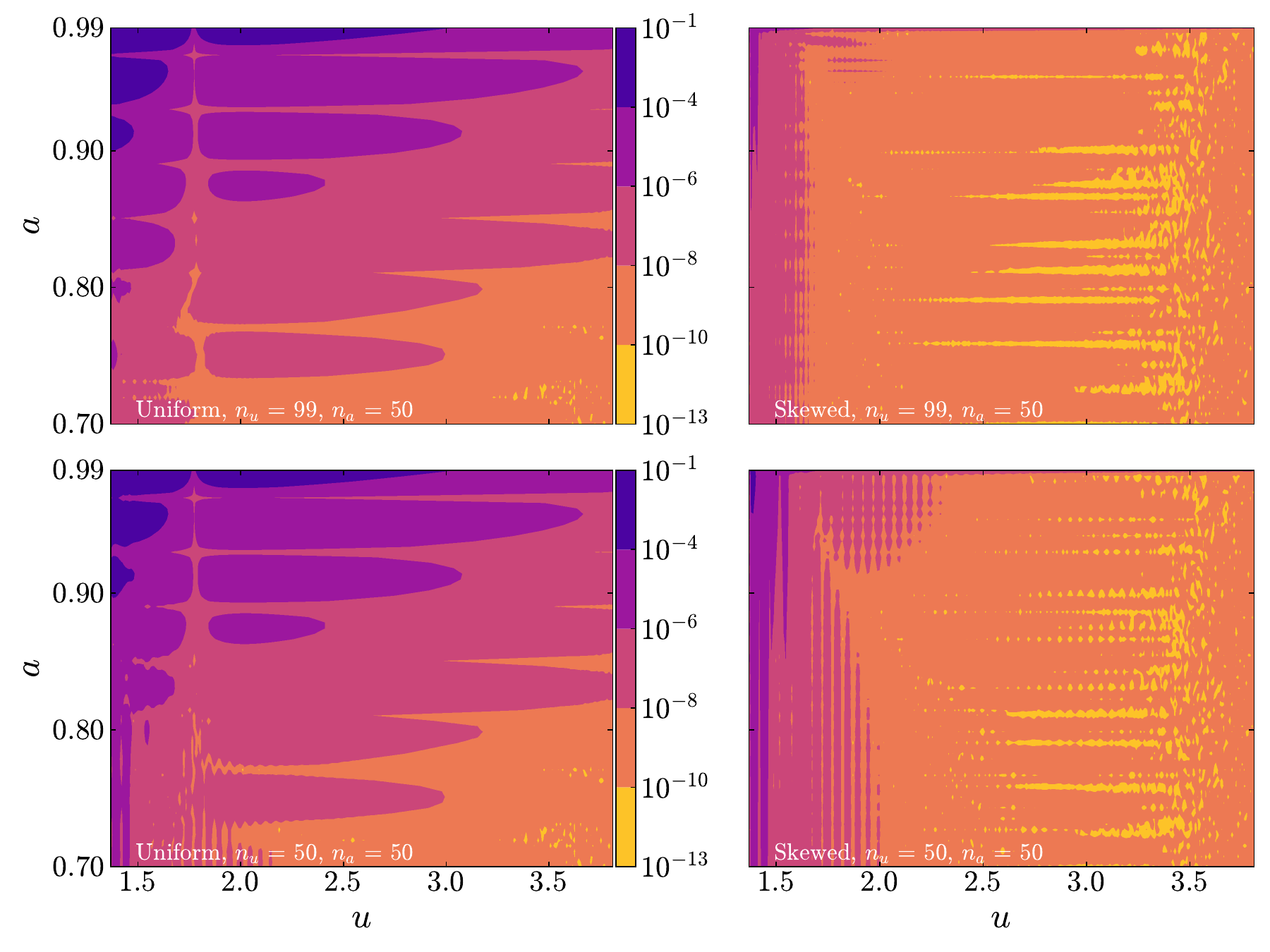}
\caption{The relative error for spline interpolated fluxes across the $(u,a)$ domain for four different interpolants trained on different down-sampled grids. The panels on the left were both made with interpolants trained on a uniform spacing in spin $\Delta a = 0.04$. The panels on the left were made with interpolants trained on a grid that used a skewed power law spacing in spin. The top panels correspond to grids with $99 \times 50$ points while the bottom panels correspond to $50 \times 50$ points.} 
\label{fig:interp_err_downsampled}
\end{figure}
Building on the comparison in Fig.~\ref{fig:interp_error_uniform_vs_skewed}, which showed the advantage of the skewed power-law spin grid at full resolution, we next investigate how far the grids can be coarsened before accuracy degrades. This step is essential because any higher-dimensional extension of the flux tables will benefit from using as few points as possible. We first halve the spin sampling, keeping $n_u = 99$ but reducing the spin grid to $n_a=50$; the resulting interpolation errors are displayed in the upper panels of Fig.~\ref{fig:interp_err_downsampled}. We then halve the separation sampling as well, adopting $n_u = 50$ together with $n_a = 50$;  these results appear in the lower panels. Even after both reductions, the non-uniform spin grid continues to control the error in the high-spin region, whereas the uniform grid shows a clear deterioration. This outcome highlights that interpolation accuracy is limited primarily by the spin grid: once the spin points are placed optimally, the logarithmically scaled separation variable $u$ already provides sufficient resolution, so further coarsening in 
$u$ has little impact on the overall error.

\begin{figure}[t]
    \centering
    \includegraphics[width=  \linewidth]{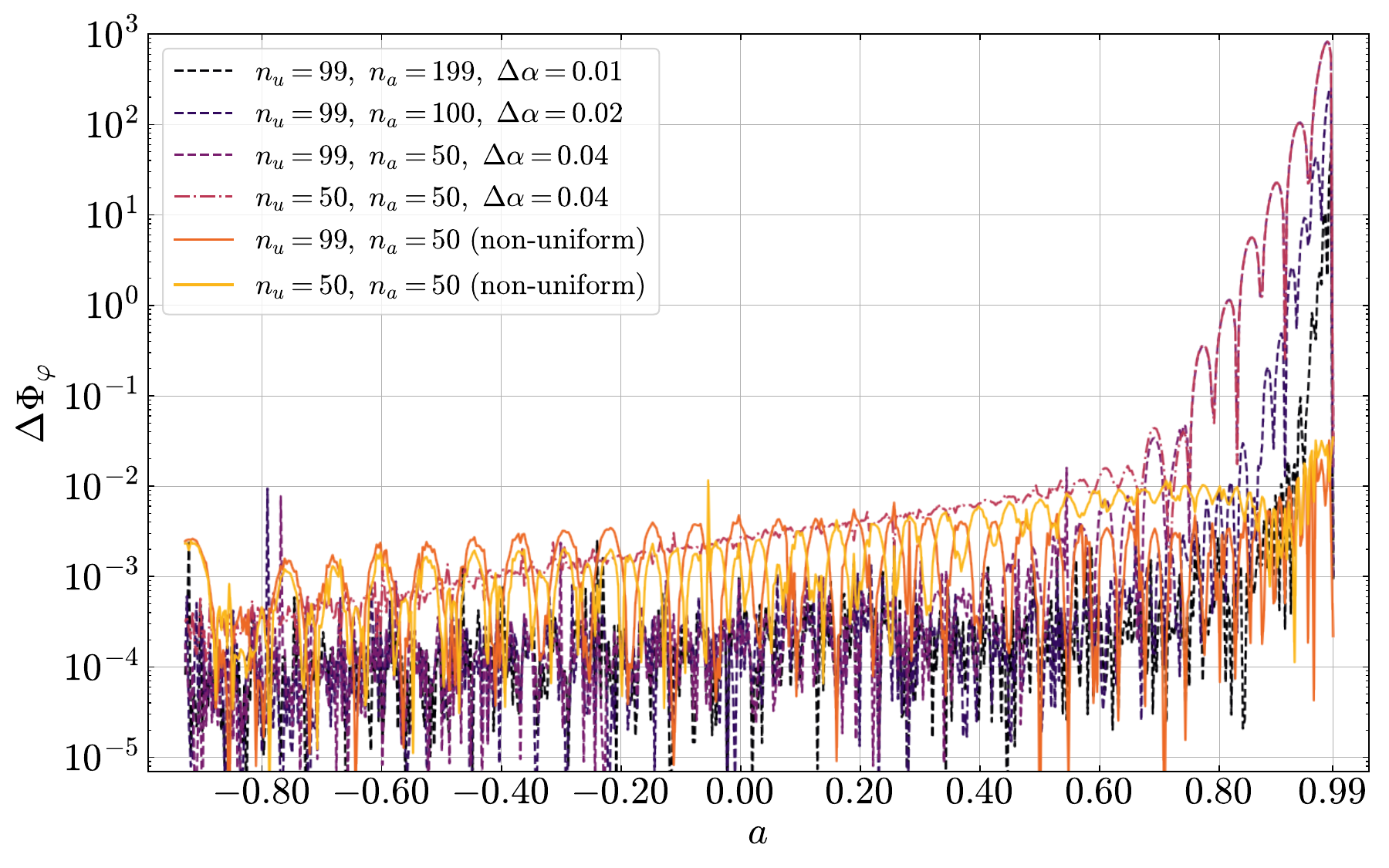}
    \caption{Accumulated orbital phase shift $\Delta\Phi_{\phi}$ over a 4-year inspiral as a function of spin $a$, for various interpolated flux grids. The masses are fixed to $M=10^6 M_{\odot}$ and $\mu=10 M_{\odot}$. The reference model uses the non-uniform, skewed power-law grid with $n_u = 99$, $n_a = 100$. Uniform grids show growing error at high spin; dips in $\Delta\Phi$ align with input spin points due to spline interpolation. %Detectable levels are typically $\gtrsim 1$ rad.
}
    \label{fig:phase_spin_spline}
\end{figure}

To assess more directly how the interpolation errors discussed above affect waveform accuracy, we construct inspiral trajectories using each of the flux grids analyzed so far. For reference, we adopt the model built from the non-uniform spin grid with $n_u = 99$ and $n_a = 100$, which—as seen in Fig.~\ref{fig:interp_error_uniform_vs_skewed}—has the lowest interpolation error across the $(u,a)$ domain, with values below $10^{-8}$ almost everywhere and below $10^{-6}$ even in the strong regions. Using this model as the baseline, we chose \gls{EMRIs} with fixed masses of $M=10^6 M_{\odot}$ and $\mu=10 M_{\odot}$ and we compute the accumulated phase shift $\Delta\Phi$ over a 4-year inspiral for each of the other grid configurations. The setup is described in detail in Sec.~\ref{sec:comparison_methods}. The 4 years accumulated phase shift data is shown in Fig.~\ref{fig:phase_spin_spline} as a function of spin parameter. Overall, the results confirm that uniform spin grids perform poorly in the prograde, high-spin region ($a \gtrsim 0.8$), whereas they may still be sufficient for retrograde or low-spin cases. We also observe an oscillatory pattern in the phase shift curves, particularly for the uniform-grid configurations. These oscillations arise from the structure of the spline interpolation: the local minima in $\Delta\Phi$ align with the input grid points in spin, where the interpolant exactly reproduces the flux data. Between these points, however, the interpolation error accumulates, resulting in phase shifts that can reach far beyond detectable levels (typically $0.1$–$1$ rad as a conservative threshold for \gls{LISA} detectability \cite{Burke:2023lno}).

\subsubsection{Error in the Chebyshev interpolation}\label{sec:chebyshev_errors}

\begin{figure}[t]
    \centering
    \includegraphics[width= \linewidth]{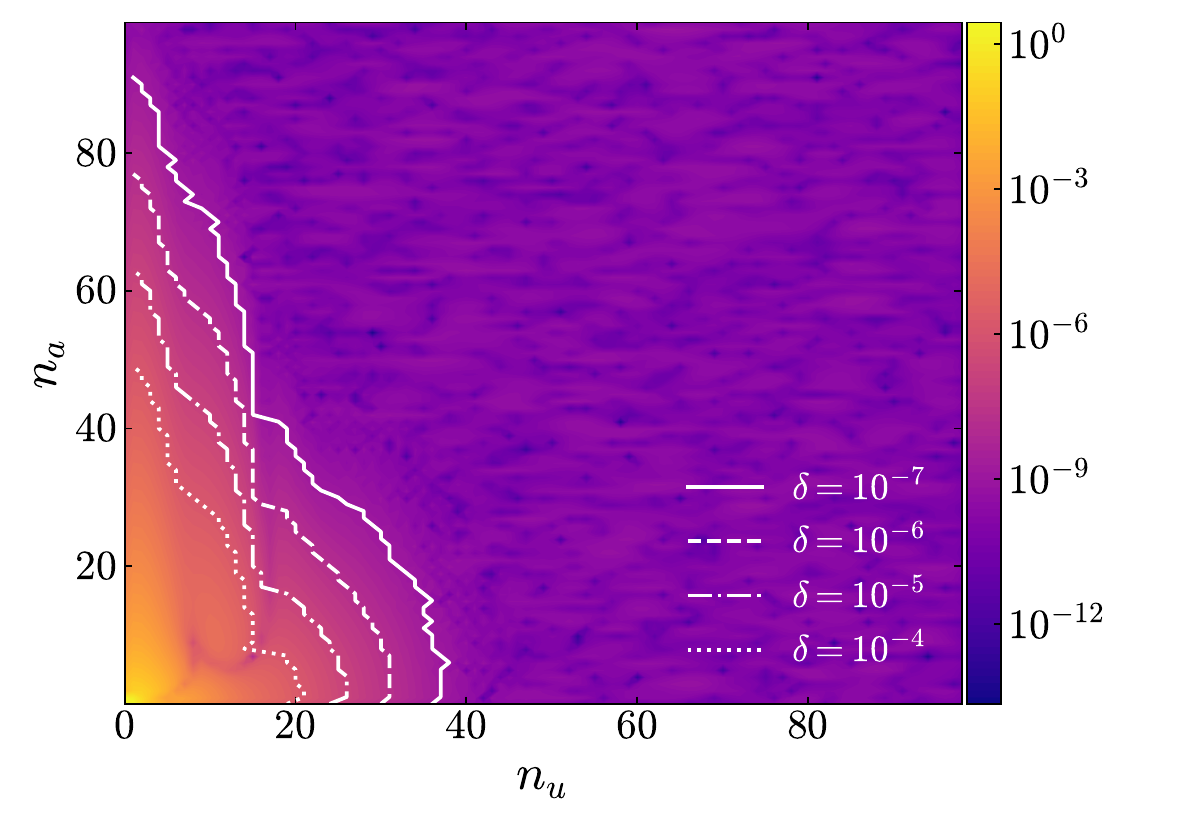}
    \caption{The magnitude of the Chebyshev coefficients relative to the largest coefficient. We also include the contours which show which coefficients are kept for different values of our relative tolerance $\delta$. We see that most coefficients contribute negligibly to the accuracy of the interpolant and so can be safely discarded.}
    \label{fig:CheybyshevCoefficients}
\end{figure}

We now investigate the effect of modelling the flux by using Chebyshev interpolants truncated to different orders to achieve a target relative error $\delta$ (Sec.~\ref{sec:eff-cheby-interp}). In Fig.~\ref{fig:CheybyshevCoefficients} we illustrate the relative magnitude of each Chebyshev coefficient compared to the largest coefficient. We see that most of the power is concentrated in the lowest order coefficients. We then plot the contours delineating the coefficients that are kept versus the coefficients that are pruned for different values of $\delta$. This demonstrates that a large number of these coefficients can be safely discarded with a negligible impact on the accuracy of the interpolant.

\begin{figure}[t]
    \centering
    \includegraphics[width= \linewidth]{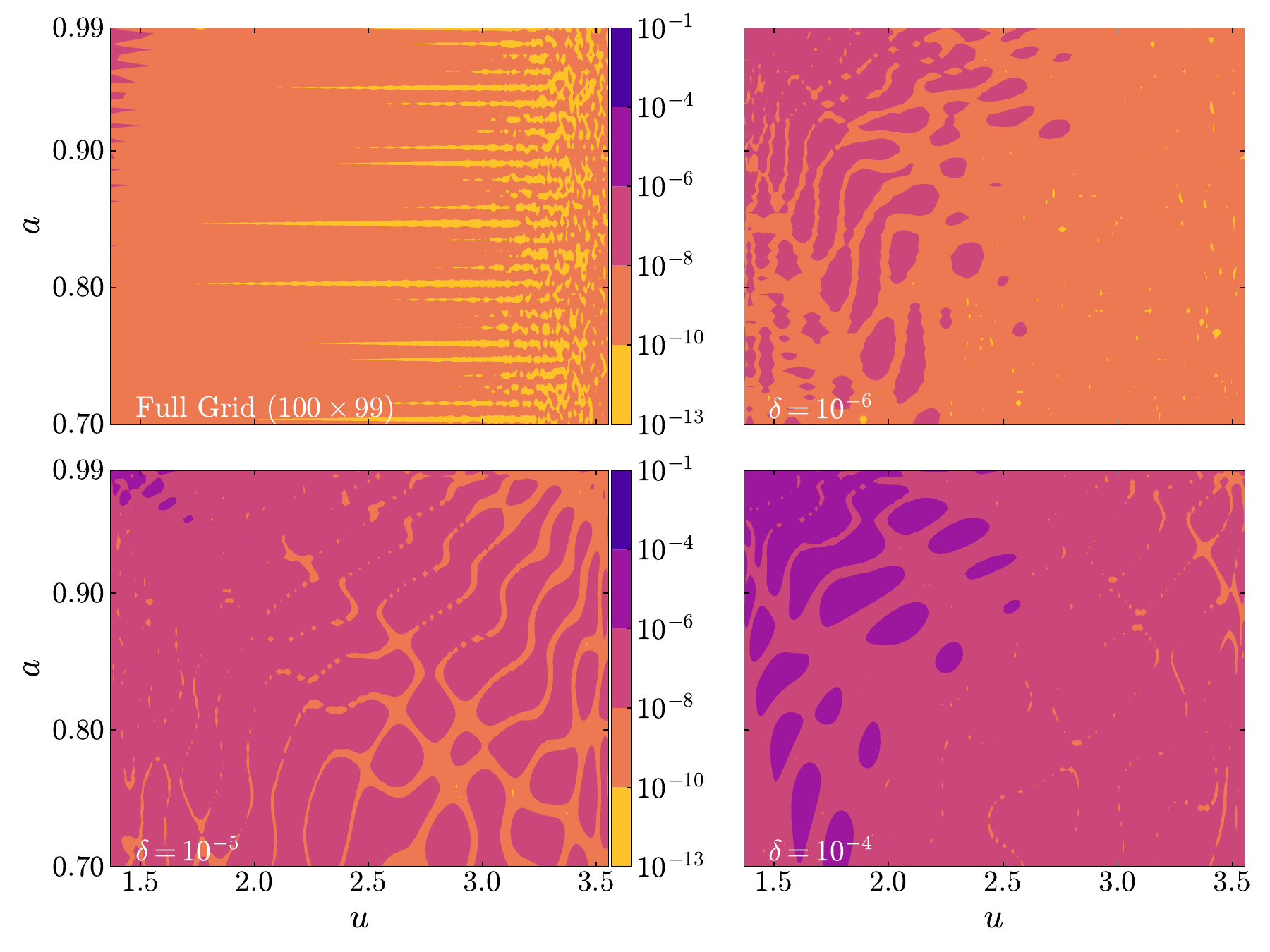}
    \caption{The the contours represent the fractional error of the different Chebyshev interpolants used in our investigation compared to an equally spaced $250 \times 450$ test grid. The largest errors are occur for high spin prograde orbits close to the last stable circular orbit.}
    \label{fig:CheybyshevFluxErrors}
\end{figure}

As in the previous section, we now test the accuracy of these interpolants by comparing to exact values on our denser test grid.
In Fig.~\ref{fig:CheybyshevFluxErrors}, we see the relative error of these interpolants across the parameter space.  As before, the errors are most prevalent for high prograde orbits close to the last stable circular orbit. While this corner of the parameter space is the most difficult to model, the rest of the parameter space is comparably well behaved. As such we expect parameter estimation biases to be largest for prograde inspirals around rapidly rotating primaries. Note that using our full $(100\times 99)$ Chebyshev interpolant (top right panel) compares favorably to using a spline with the same number of points using either a uniform or skewed grid as shown in Fig.~\ref{fig:interp_error_uniform_vs_skewed}, especially for low values of $u$ which are close to the separatrix. This is due both to the exponential convergence of Chebyshev interpolation and the natural concentration of Chebyshev nodes at high spins and deep in the strong field. 

In Table~\ref{tab:Chebysehv_Interps}, we list both the target relative error $\delta$ and the maximum relative error as measured against the uniform test grid. As we can see, $\delta$ proves to be a reasonable upper bound for the measured relative error of the interpolant.
The number of grid points necessary to achieve the target $\delta$ is estimated by the length and largest value of $n_{y,i}$ (Sec.~\ref{sec:eff-cheby-interp}). This shows that one can get away with far fewer points in $u$ than in $a$.

In order to investigate the effect these errors would have on the speed and accuracy of an inspiral, we first chose a typical EMRI system with primary mass of $10^6 M_\odot$ and secondary mass of $10 M_\odot$ and vary the size of $\delta$. 

In order to engineer a worst case scenario for the interpolation error within our spin domain, $a\in [-0.99, 0.99]$, we choose a large prograde spin of $a = 0.9899$, close to the upper boundary and intentionally off the interpolation grid, since interpolants typically exhibit their largest errors away from the grid points. We start the inspiral at a separation of $p_0 = 10$, so that the compact object plunges just before the end of the 4 year mission lifespan of \gls{LISA}. We run this same trajectory 1000 times and report the mean and standard deviation of the time taken for the trajectory calculation on an Apple M1 Max. For reference, using a spline interpolant takes $3.25 \pm 0.52$ms \footnote{We find that the timings for bicubic spline interpolants is insensitive to the number of grid points which is consistent with the evaluation time scaling as $\mathcal{O}(1)$.}and using the 5PN expressions for the fluxes takes $9.26 \pm 0.69$ms. As such, using our Chebyshev interpolation code with $\delta \geq 10^{-5}$ would be as fast as using the analytic PN expressions (column 4 of Tab.~\ref{tab:Chebysehv_Interps}). 

We also report the azimuthal dephasings against the full Chebyshev interpolant. From this simple analysis, we see that for $\delta \leq q$, the dephasing is $\lesssim 0.1$ radians, while larger values of $\delta$ show significantly more  dephasing.
We also report the mismatch with the full Chebyshev interpolant using a flat \gls{PSD}. We notice that for $\delta \leq q$ the mismatches are $\leq 10^{-3}$, but these mismatches become significantly worse when $\delta > q$. These results further support the hypothesis that using $\delta \sim q$ is a reasonable target precision to avoid significant bias. 

\begin{table*}[]
    \centering
    \renewcommand{\arraystretch}{1.2}
    \setlength{\tabcolsep}{8pt}
    \begin{tabular}{|c|c|c|c|c|c|}
    \hline
    $\delta$ & $n_u \times n_a$  & Max Rel. Err. & Traj. Time [ms] $$ & $\Delta \Phi_\phi$ [rad] & $\mathcal{M}$  \\
    \hline \hline
     $8 \times 10^{-8} $ & $99 \times 100$   & $4.14 \times 10^{-8}$ & $169.9 \pm 2.5$ & - & - \\
    \hline
    $10^{-7}$   & $38 \times 92$ & $7.95 \times 10^{-8}$ & $17.9 \pm 0.9$ &  $7.03 \times 10^{-4}$ & $1.8 \times 10^{-5}$ \\
    \hline
    $10^{-6}$   & $31 \times 78$ & $1.02 \times 10^{-6}$ & $14.24 \pm 1.3$ & $1.03 \times 10^{-2}$ & $4.9 \times 10^{-5}$ \\
    \hline
    $10^{-5}$   & $26 \times 64$ & $9.35 \times 10^{-6}$ & $10.11 \pm 0.13$  & $0.103$ & $7.1 \times 10^{-4}$ \\
    \hline
    $10^{-4}$   & $21 \times 50$ & $7.62 \times 10^{-5}$ & $6.98 \pm 0.66$ & $2.808$ & $0.315$ \\
    \hline
    $10^{-3}$   & $15 \times 37$ & $7.55 \times 10^{-4}$& $4.89 \pm 0.09$ & $15.288$ & $0.516$  \\
    \hline
    \end{tabular}
    \caption{A list of the different Chebyshev interpolants that we use in our investigation along with their associated estimated max relative errors $\delta$ and measured max relative error compared to an independent grid. The number of grid points is estimated by the length and largest value of $n_{y,i}$ used to achieve $\delta$. We also run an inspiral with $M = 10^6 M_\odot$, $\mu$ = $10 M_\odot$, $r_0 =10$, $a = 0.9899$, 1000 times and report the mean and standard deviation of the runtime for the trajectory. For reference, using a spline interpolant takes $3.25 \pm 0.52$ms and using the 5PN expressions for the fluxes takes $9.26 \pm 0.69$ms. We also report the azimuthal dephasings and flat \gls{PSD} mismatch when compared with the full Chebyshev interpolant.}
    \label{tab:Chebysehv_Interps}
\end{table*}

\begin{figure}[t]
    \centering
    \includegraphics[width=  \linewidth]{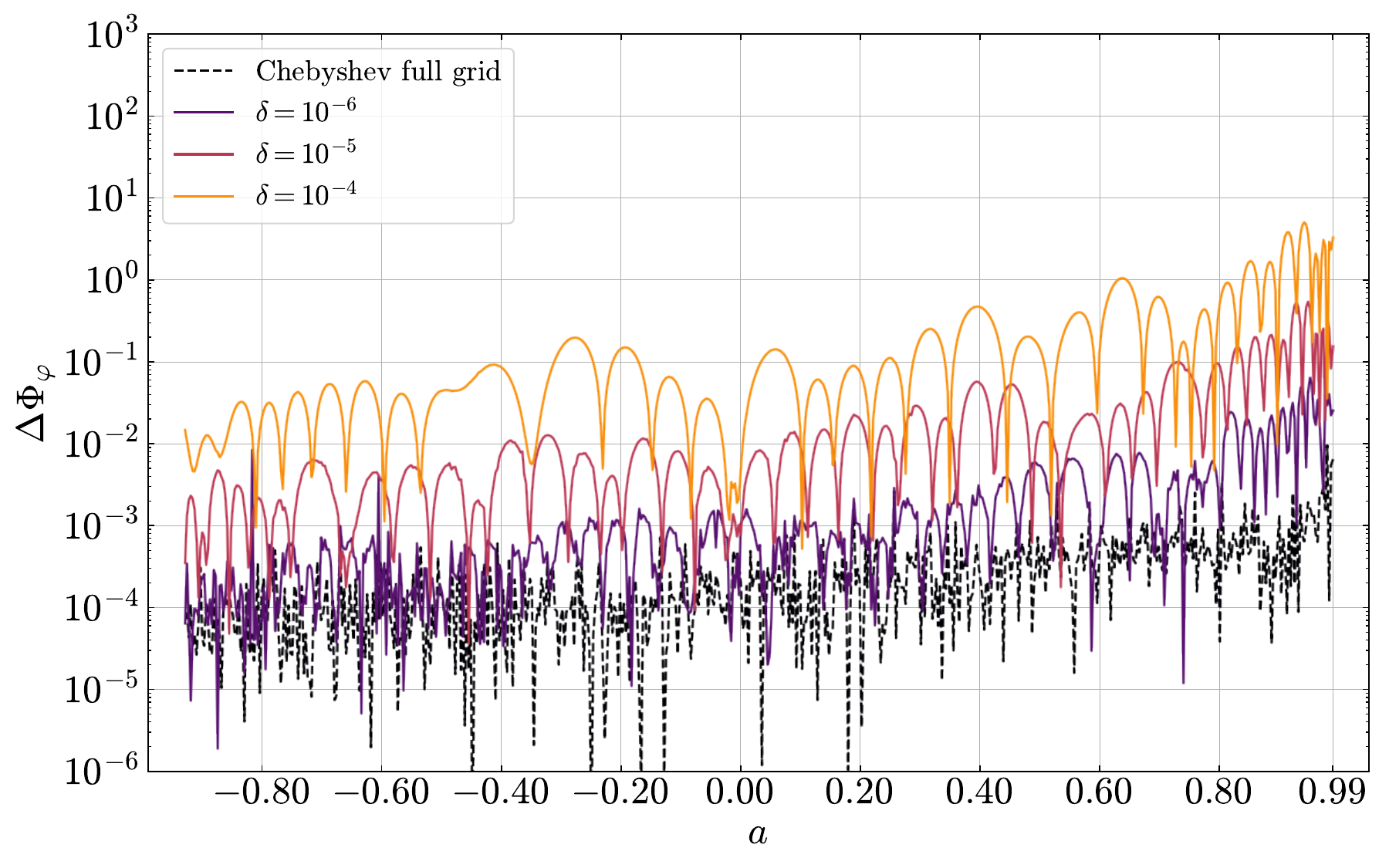}
    \caption{The accumulated difference in the orbital phase for, when one changes the $\delta$ for Chebyshev interpolant for the gravitational wave flux for different values of primary spin $a$. The reference trajectory model uses a spline interpolant trained on the $99 \times 100$ non-uniform, skewed power-law grid, which we use as a reference. }
    \label{fig:ChebyshevDephasings}
\end{figure}

To ensure that this is a reasonable test case, we examine the azimuthal dephasing against our reference trajectory of the spline case with the flux trained on a $99 \times 100$ skewed non-uniform grid. We vary the value of $a$ for a selection of different Chebyshev interpolants over 4 years of inspiral to plunge. From Fig.~\ref{fig:ChebyshevDephasings}, one sees that the largest dephasing occur for large prograde spins and that our interpolants perform better elsewhere in the parameter space. 

Additionally, we see that using $\delta \leq q$ results in dephasings that are consistently $< 1$ radian, while this is not guaranteed for larger values of $\delta$. 

Moreover, the Chebyshev interpolants show a flatter dephasing trend---aside from small oscillations at grid points---due to their global error control, unlike the spline case, which exhibits a steep phase error rise in the high-spin regime from its localized nature.

\begin{figure}[t]
    \centering
    \includegraphics[width=  \linewidth]{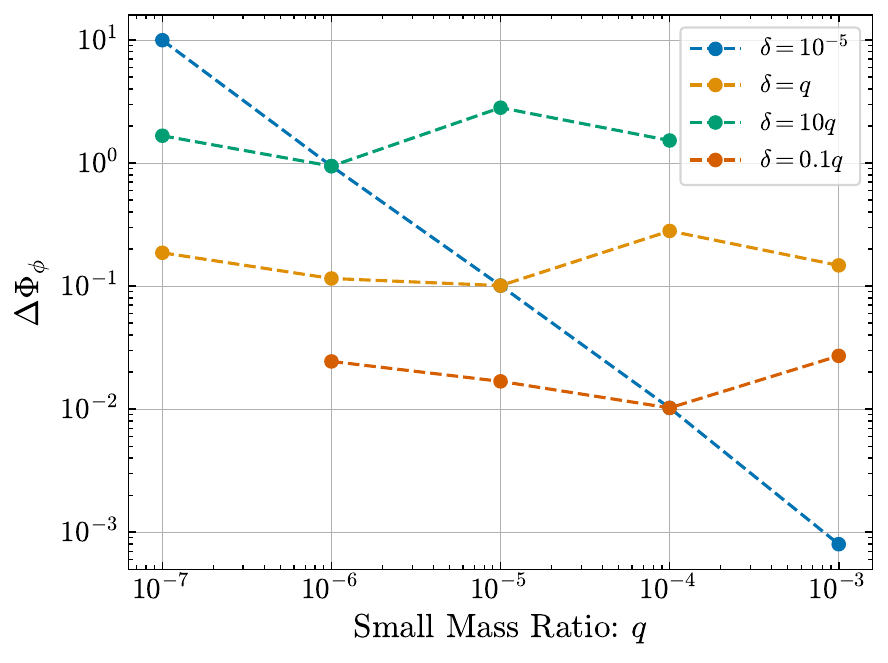}
    \caption{
    The azimuthal dephasing for different interpolant models compared to the full $99 \times 100$  Chebyshev interpolant for different values of $q$. For each inspiral $a = 0.9899$ and $r_0 = 10$ and is evolved until plunge. The green/yellow/red dashed-dotted curves represent the accumulated dephasing $\Delta \Phi_\phi$ for Chebyshev interpolants with truncation errors $\delta = \{10q,q,0.1q\}$. The dashed-dotted blue curve plots the accumulated dephasing for a fixed truncation error $\delta = 10^{-5}$. In order to obtain dephasings that are consistently $\lesssim 1$ radian for all mass ratios, one should use the target relatively accuracy $\delta = q$.}
    \label{fig:CheybyshevDeltaVsq}
\end{figure}

\vspace{10pt}
Finally, to demonstrate how these results would vary with different mass ratios we measure the dephasing between the full Chebyshev interpolant and interpolants with different values of $\delta$ while varying the mass ratio. Each inspiral has $M = 10^6 M_\odot$, $a = 0.9899$, and starts at radial coordinate $p_0 = 10$ that is then evolved until plunge. While some of the resulting systems might take hundreds of years to plunge and/or have a subsolar mass secondary, we are only concerned with ensuring that the systems always evolve through the same values of $p$ in order to isolate how the dephasing scales with $q$. The results of this analysis are displayed in Fig.~\ref{fig:CheybyshevDeltaVsq}. 

From figure \ref{fig:ChebyshevDephasings}, we understood that a Chebyshev truncation error of $\delta = 10^{-5}$ was sufficient to keep the dephasing $\Delta \Phi_{\phi} \lesssim 1$ for a four-year inspiral with mass-ratio of $q = 10^{-5}$. Similarly, from Tab.\ref{tab:Chebysehv_Interps}, the row corresponding to $\delta = 10^{-5}$ yields $\Delta \Phi_\phi \sim 0.1$ and $\mathcal{M}\lesssim10^{-3}$ for a spin parameter $a = 0.9899$. As demonstrated in both Fig.~\ref{fig:ChebyshevDephasings} and Tab.~\ref{tab:Chebysehv_Interps}, increasing the truncation error $\delta$ significantly degrades the accumulated dephasing and mismatches between the approximate and true models. From Fig.~\ref{fig:CheybyshevDeltaVsq}, we found that the optimum truncation error was $\delta \sim q$. By setting $\delta \sim q$, the resulting interpolant consistently resulted in dephasings $\Delta\Phi_{\phi} \sim 0.1$, whilst still retaining reasonable evaluation times. Setting $\delta = 10~q$ results in a dephasing that is consistently $>1$ radian and setting $\delta = 0.1 q$ gives more accuracy than would be required. We will more thoroughly test this hypothesis in the next section using a full Bayesian analysis on more realistic EMRI signals one would expect to see with \gls{LISA}.

\subsubsection{Waveform fidelity and Bayesian Inference} \label{sec:interp_error_mcmc}

Building on the results from the flux error and phase shift analyses from Sec.~\ref{sec:interpolation_error} and Sec.\ref{sec:chebyshev_errors}, we now evaluate the impact of interpolation-driven inspiral errors on waveform fidelity. To do this, as the first step we compute waveform mismatches relative to a high-accuracy reference model, using a selected subset of inspiral models discussed previously. \\
From our assessment of the flux error in Fig.~\ref{fig:CheybyshevFluxErrors}, we now use the full $99 \times 100$ Chebyshev grid inspiral model as the reference, as it demonstrated the lowest interpolation error across the entire $(u,a)$ domain.\\
We construct waveforms using five other inspiral trajectories: 
\begin{itemize}
    \item Spline interpolant on a non-uniform grid of $n_u = 99$ by $n_a = 100$ points
    \item Spline interpolant on a non-uniform grid of $n_u = 99$ by $n_a = 50$ points
    \item Spline interpolant on a non-uniform grid of $n_u = 99$ by $n_a = 50$ points
    \item Chebyshev interpolant with $\delta = 10^{-5}$ (effective grid of $n_u = 26$ by $n_u = 64$ points )
    \item Chebyshev interpolant with $\delta = 10^{-4}$ (effective grid of $n_u = 21$ by $n_u = 50$ points )
\end{itemize}
We exclude the remaining inspiral models from this comparison, as the earlier phase shift results already showed that their errors are too large to be viable for parameter estimation and would inevitably lead to significant biases.

\begin{figure}[t]
    \centering
    \includegraphics[width= \linewidth]{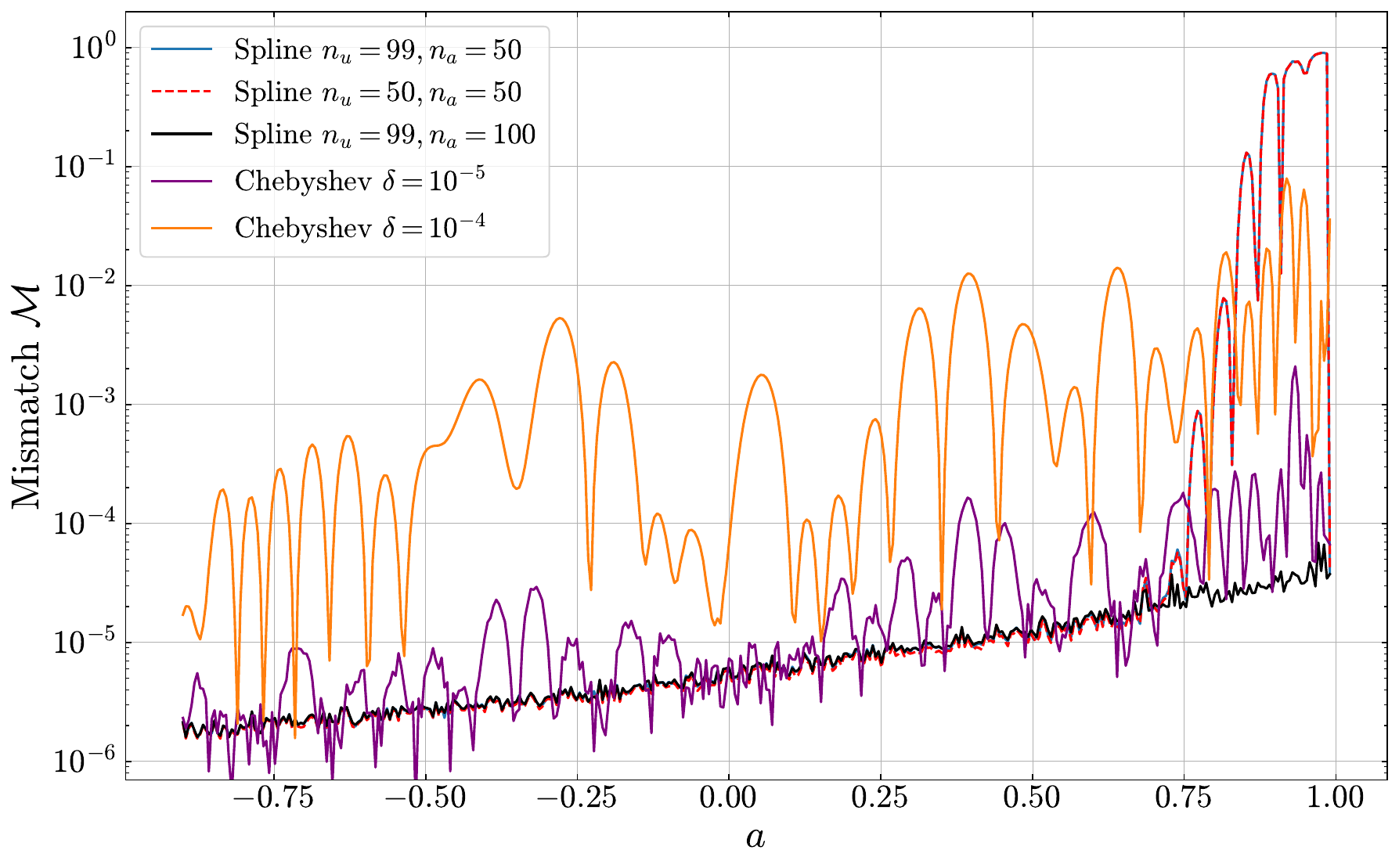}
    \caption{The mismatch $\mathcal{M}$ between the the full $99 \times 100$ Chebyshev model as a function of spin $a$ for series of different models. The blue, dashed-red and black curves represent the mismatch from using spline interpolants trained on non-uniform grids with different resolutions. The purple and orange curves represent the mismatch from using Chebyshev interpolants with $\delta$ set to either $10^{-5}$ or $10^{-4}$. In all cases, the $\mathcal{M}$ is greatest for large prograde spins. }
    \label{fig:mismatch_interp}
\end{figure}

We compute the mismatches between waveforms, with $p_0$ chosen to ensure a time-to-plunge of 4 years with fixed masses of $M = 10^6,M_\odot$ and $\mu = 10,M_\odot$. The mismatch results are shown in Fig.~\ref{fig:mismatch_interp}, plotted as a function of spin $a$. Since all waveform models share the same mode amplitudes and differ only in the inspiral trajectory, any observed mismatch reflects the impact of flux-induced trajectory errors. As expected, the full-resolution spline model with the non-uniform spin grid achieves the lowest mismatch across the spin range, remaining below $10^{-4}$ throughout.

An oscillatory pattern is again visible, consistent with the behavior seen in the phase shift plots---minima in the mismatch align with the input grid points. The spline-based models show a clear increase in mismatch at high spin, whereas the Chebyshev-based models exhibit a flatter trend. This reflects the earlier flux error maps, where Chebyshev grids distributed error more uniformly, while spline grids showed larger errors concentrated in the high-spin region.

We also note that using Chebyshev interpolant with $\delta = 10^{-4}$ does not quite meet the requirement of $\mathcal{M} \lesssim 10^{-2}$ everywhere in the parameter space, but $\delta = 10^{-5}$ does. This give additional credence to the claim that $\delta \sim q$ may be sufficient for parameter estimation.

Next, we perform a parameter estimation bias analysis using full \gls{MCMC} sampling, comparing approximate waveform models against a high-accuracy reference. As before, we adopt the full-resolution Chebyshev grid model as the reference, which serves as the most accurate inspiral trajectory in our study. For this analysis, we will use second-generation TDI variables with assuming a static arm-link configuration of LISA (equal and constant arm-lengths). We will use \texttt{fastlisaresponse}, a GPU accelerated time-domain response function that accounts for the propagated gravitational radiation onto the LISA instrument\cite{Katz_2022}. Our sampling techniques were described earlier in Sec.\ref{sec:comparison_methods}.

\begin{table*}[t]
    \centering
    \renewcommand{\arraystretch}{1.2}
    \setlength{\tabcolsep}{8pt}
    \begin{tabular}{|c|cc|cc|cc|}
        \hline\hline
        Mass-ratio $q$ & \multicolumn{2}{c|}{$\delta = 0.1q$} & \multicolumn{2}{c|}{$\delta = q$} & \multicolumn{2}{c|}{$\delta = 10q$} \\
        \hline
        &  $\Delta \Phi$ [rad] & $\mathcal{M}$ 
        & $\Delta \Phi$ [rad] & $\mathcal{M}$ 
        & $\Delta \Phi$ [rad] & $\mathcal{M}$ \\
        \hline
        $10^{-4}$ & $2.34\times 10^{-4} $ &  $2.13\times10^{-5} $ & 1.280 & 0.262 & 5.386 & 0.958 \\
        $10^{-5}$ & 0.0177 & $9.18 \times 10^{-5} $ & 0.495 & $5.66\times 10^{-2}$ & 0.958 & 0.377 \\
        $10^{-6}$ & $2.92 \times 10^{-3} $ & $2.30 \times 10^{-6}$ & $3.29\times 10^{-2} $ & $2.05 \times 10^{-4}$ & 0.614 & 0.108 \\
        \hline\hline
    \end{tabular}
    \caption{Dephasing $\Delta \Phi$ (in radians) and mismatch calculations $\mathcal{M}$ for different mass-ratios $q\in\{10^{-4},10^{-5},10^{-6}\}$ using trajectories built with Chebyshev interpolants of different $\delta \in \{0.1,1,10\}q$. All trajectories have a primary mass $M = 10^6M_\odot$ and an SNR of 100. The mismatches and dephasings here are computed at the true parameters, comparing the true waveform model (full-resolution Chebyshev interpolant) to the approximate model using an interpolant with maximum error set by the table.}
    \label{tab:dephasing_mismatch}
\end{table*}

We focus only on the Chebyshev-based models for two reasons. First, their mismatches lie in an intermediate regime---not clearly negligible, yet not large enough to be definitively detectable---making them ideal for probing potential biases near the detection threshold. Second, our Chebyshev approach offers direct control over global interpolation accuracy via the parameter $\delta$, allowing the flux model to be systematically tuned to the desired precision based on system properties such as the mass ratio $q$.

In our PE runs, for our approximate model, we will use three approximate Chebyshev interpolants with maximum relative errors set depending on the mass-ratio of the system $q$. For the analysis, we will fix the primary mass $M = 10^{6}M_{\odot}$ but change the mass-ratio $q \in \{10^{-6},10^{-5},10^{-4}\}$ and fix $p_{0}$ such that we observe a $T = 4$ year long waveform sampled with cadence $\Delta t = 5\,$ seconds up until the plunge. For each waveform model, we fix the full mode structure $A_{lm}$ with $l\in [2,10]$ and $m \in [-l, \ldots, 0,\ldots, l]$. This is so that biases recovered can only be due to dephasing between waveform models (due to inaccurate trajectories entirely driven by the fluxes) rather than mismatches in amplitude. The corresponding phase shifts and mismatches for each approximate model used in our \gls{MCMC} runs are summarized in Table~\ref{tab:dephasing_mismatch}. For all waveform models, we will fix the dimensionless spin parameter $a = 0.9264$ as noted earlier and with the same angular parameters described in Tab.I in \cite{chapman2025fast}. 

In our analysis, we will fix the SNR = 100 by tuning the luminosity distance to the source, which is a conservative criterion for the mass-ratio cases $q = 10^{-5}$ and $q = 10^{-6}$ (see Fig.17 in \cite{chapman2025fast}). As demonstrated in ~\cite{Cutler:2007mi, Miller2005, Flanagan:1997kp}, biases as a result from waveform mis-modeling are SNR independent, whereas the precision in which parameters are measured is inversely proportional to the SNR. For large SNR sources, with tight parameter uncertainties, parameter biases become more apparent. A criterion popularized in Cutler-Valisneri~\cite{Cutler:2007mi} suggests that waveform modelling errors are only suitable for parameter estimation if and only if the recovered (biased) best-fit parameters are contained within the $1\sigma$ regime of the posterior. This is to ensure that parameter biases are consistent with ($1\sigma$) statistical fluctuations induced via instrumental noise realizations, assuming that the underlying noise model has been properly described (see ~\cite{burke:2025bun} for further discussion). For precisely this reason, we will neglect additive noise in our likelihood calculations but retain the PSD in \eqref{eq:inner_product} and \eqref{eq:likelihood} to properly account for uncertainty in parameter measurements due to the presence of noise. Recovered parameters that deviate from the true parameters are \emph{only} a result of waveform mismodelling errors. From here-on, we will adopt a slightly more general form of the Cutler-Valisneri criterion~\cite{Cutler:2007mi}. We will regard a waveform model as suitable for parameter estimation if and only if the true parameter lies within the 68\% Highest Posterior Density Interval (HPDI) of the approximate posterior distribution. This is discussed at great length in ~\cite{Burke:2023lno}, Sec. III D.

\begin{figure*}
    \includegraphics[width =\linewidth]{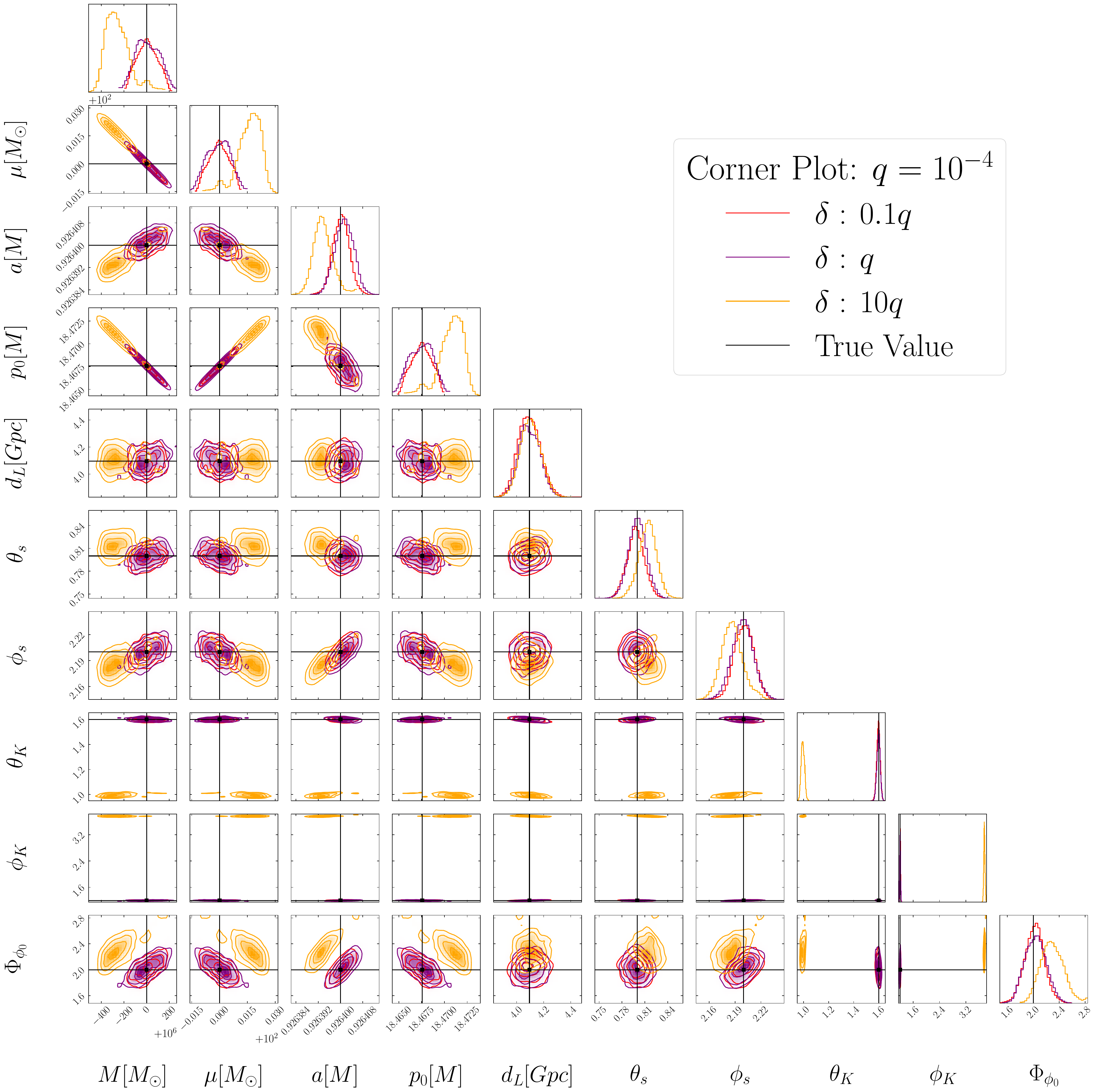}
    \caption{Plot of posterior distributions generated via performing inference on an injected waveform (Full Chebyshev) but recovered with waveform models using approximate Chebyshev interpolants with maximum errors set to $\delta \in\{0.1q,q,10q\}$ for the blue, red and purple posterior respectively. Here we fix the mass-ratio $q = 10^{-4}$ with primary mass $M = 10^{6}M_{\odot}$ and consider a source with SNR = 100. We find that a relative error of order $q = 10^{-4}$ is suitable for parameter estimation for this particular configuration.}
    \label{fig:MCMC_result_1eneg4}

\end{figure*}

\begin{figure*}
\includegraphics[width =\linewidth]{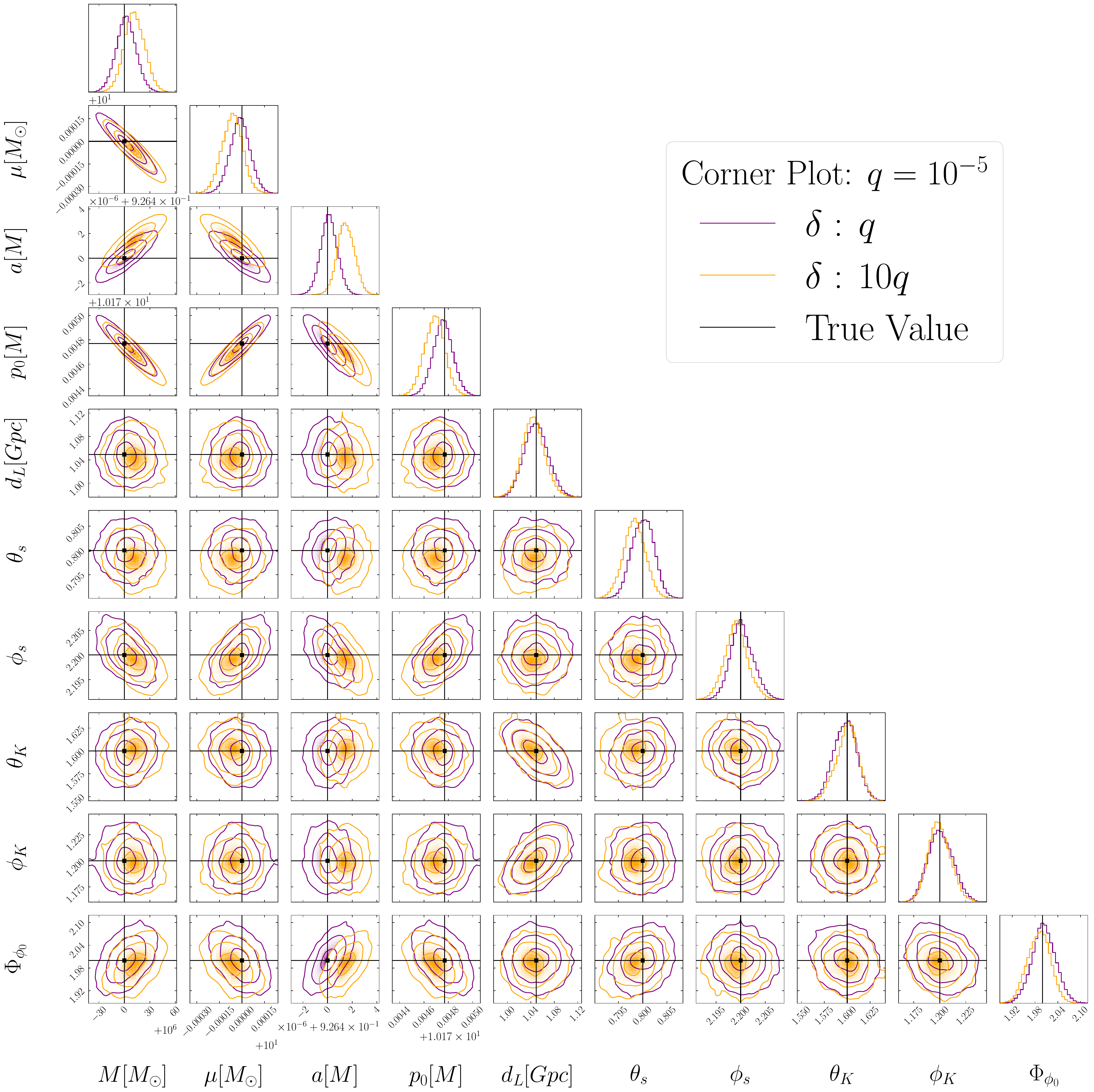}
\caption{The same set up as Fig.~\ref{fig:MCMC_result_1eneg4} except with $q = 10^{-5}$. Similarly, we find that a relative error of order $q = 10^{-5}$ is sufficient for parameter estimation.}
\label{fig:MCMC_result_1eneg5}

\end{figure*}

\begin{figure*}
\includegraphics[width =\linewidth]{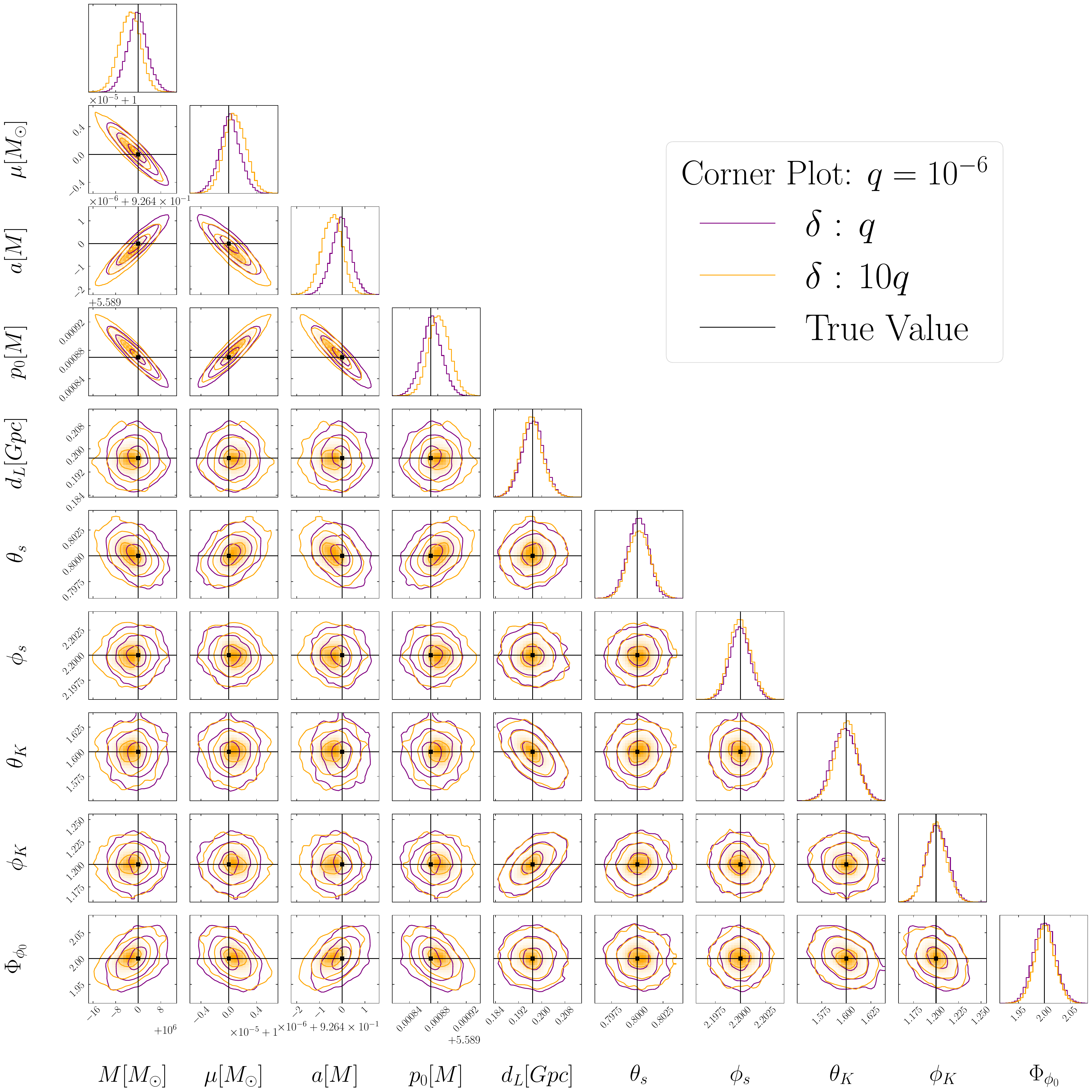}
\caption{The same set up as Fig.~\ref{fig:MCMC_result_1eneg4} and \ref{fig:MCMC_result_1eneg5} except with $q = 10^{-6}$. Interestingly, we find that a relative error of order $10q = 10^{-5}$ is sufficient for parameter estimation.}
\label{fig:MCMC_result_1eneg6}

\end{figure*}

For each of the mass-ratios $q\in\{10^{-4},10^{-5},10^{-6}\}$ we provide corner plots in Figures \ref{fig:MCMC_result_1eneg4}, \ref{fig:MCMC_result_1eneg5} and \ref{fig:MCMC_result_1eneg6} respectively. In each case, the blue, red, and purple posterior reflects model waveforms with less accuracy than the injected waveform with global interpolant truncation error set to $\delta \in \{0.1q, q, 10q\}$, respectively.

Starting with mass-ratio $q = 10^{-4}$ case presented in Figure~\ref{fig:MCMC_result_1eneg4}, we observe that waveform models with flat errors of $10^{-5}$ and $10^{-4}$ are suitable for parameter estimation. In each case, the true parameters lie within the approximate 68\% HPDI. The model with error $10^{-3}$ is clearly unsuitable for parameter estimation, since almost all of the parameters lie outside the 68\% credible interval. We remark here that for parameter estimation purposes we would require a minimum error on the interpolants on the order of $q = 10^{-4}$. For search purposes, however, we could get away with an error $\sim 10q = 10^{-3}$ since we are more interested in finding regions of parameter space that contain the true parameters of the source. We find similar results in Fig.~\ref{fig:MCMC_result_1eneg5} and \ref{fig:MCMC_result_1eneg6}. 
We found for both these simulations that a relative error of order $q$ was sufficient for parameter estimation purposes. This is consistent with Fig.~\ref{fig:ChebyshevDephasings}, demonstrating that the global error set of the interpolant must be no larger than the mass-ratio of the EMRI system being generated online. 

In conclusion, for circular orbits in Kerr spacetime, we believe that a maximum error on the interpolants should be set to $\sim q$ for parameter estimation and potentially $\gtrsim10q$ for search purposes. 

\section{Conclusion}

In this paper, we have systematically investigated two key sources of error in fast relativistic EMRI waveform models: mode-sum truncation in radiation-reaction fluxes and interpolation inaccuracies arising from fast-generation architectures such as FastEMRIWaveforms \cite{chua2021rapid,katz2021fast,Speri:2023jte,chapman2025fast}. While adiabatic waveform models are built from fully relativistic data using black hole perturbation theory, we demonstrate that these models may still carry significant hidden systematics when used in practice for data analysis.

First, in Sec.~\ref{sec:truncation_error} we showed that truncating the angular mode sum at a representative truncation level $\ell_{\max} = 10$ introduces flux errors that can accumulate into orbital phase shifts on the order of several radians over a typical 4-year \gls{LISA} observation. 
We quantified the impact of this truncation error on parameter recovery through a full Bayesian inference study. For two-year long EMRI observations with $M = 10^{6}M_{\odot}$, mass-ratio $q = 10^{-5}$ and with even non-extremal primary spin $a = 0.9$, we demonstrated that truncating the fluxes to $\ell_{\rm max} = 10$ induces severe biases on parameters when compared to a truthful model at $\ell_{\text{max}} = 30$. However, at $\ell_{\rm max} = 20$, even at very high SNRs of $\sim 160$ we observed no statistically significant biases on the parameters. While $\ell_{\max}=20$ is sufficient for the $a=0.9$ case studied in the PE analysis, higher-spin systems may require larger mode sums, as supported by the mode and phase-error behavior shown in Figs.~\ref{fig:flux_mode} and \ref{fig:flux-traj-lmodes}; we therefore recommend $\ell_{\max}\gtrsim 30$ as a safer choice for parameter estimation in Kerr circular–equatorial systems.

This result highlights the importance of carefully handling flux truncation errors when constructing accurate EMRI waveforms. Although our analysis focuses on quasi-circular equatorial orbits, similar considerations apply to generic (eccentric and inclined) inspirals, where the flux involves additional summations over radial ($n$) and polar ($k$) mode numbers \cite{Drasco:2005is,Hughes:2005qb,Hughes:2021exa}. In those cases, the mode structure is more complex, and the flux contributions do not exhibit a simple monotonic falloff with increasing $n$ or $k$. Truncating too early may therefore not only reduce the \textit{precision} of the flux but also compromise its \textit{accuracy}, potentially leading to even larger waveform errors. This underscores the need for mode truncation choices to be physically justified across all degrees of freedom in the inspiral.

Second, in Sec.~\ref{sec:interpolation_error} we assessed how interpolation of precomputed flux data can introduce additional error in waveform generation. We analyzed both cubic spline and Chebyshev interpolation methods across different grid structures. In Sec.~\ref{sec:spline_errors} we showed that while spline interpolation on uniform grids leads to large errors in the strong-field, high-spin regime, adapting to a non-uniform, spin-skewed grid significantly improves performance. 

Furthermore, in Sec.~\ref{sec:eff-cheby-interp} we introduced an efficient Chebyshev interpolation scheme that allows explicit control over maximum global relative error through a tunable parameter $\delta$. In Sec.~\ref{sec:chebyshev_errors}, we showed that the spectral convergence of the method allows us to achieve our desired global accuracy of $\delta = 10^{-6}$ while tiling the 2-dimensional parameter space of circular Kerr orbits with only $31 \times 78$ points. These savings will become even more important when considers the higher dimensional parameter spaces associated with eccentricity \cite{Glampedakis:2002ya,Lynch:2021ogr,chapman2025fast} and orbital inclination \cite{Hughes:2001jr,Lynch:2023gpu}. As such, Chebyshev interpolation can make it significantly easier to tile the 4-dimensional parameter space of generic Kerr inspirals\cite{Drasco:2005kz,Hughes:2005qb,Hughes:2021exa, Drummond:2023wqc,Lynch:2024ohd,Lynch:2024hco}. We also demonstrated trajectory calculation times comparable to using the 5PN expressions for the fluxes. However, it remains as future work to demonstrate that such fast timings can be maintained for higher dimensional Chebyshev interpolants.

Finally, in Sec.\ref{sec:interp_error_mcmc}, we use both waveform mismatches and Bayesian inference studies to inform a practical guideline for interpolating fluxes: for reliable EMRI parameter estimation, the relative flux interpolation error should be kept below the mass ratio of the system, i.e. $\delta \lesssim q$. For detection-level searches, errors up to $\sim 10q$ may still be acceptable, but will begin to introduce measurable biases in the recovered parameters. 

In the Chebyshev case, the ability to control the relative error via a tunable parameter $\delta$ allows us to generalize this conclusion beyond interpolation. Since waveform bias is ultimately driven by the flux error regardless of its origin, our results imply a more general criterion: for quasi-circular equatorial orbits, waveform accuracy is preserved as long as the total relative flux error---whether due to interpolation, numerical inaccuracy, or modeling approximations---remains below the mass ratio, $\delta \lesssim q$. This makes $\delta$ a practical diagnostic threshold for waveform reliability. As a conservative guideline, we can set an upper limit by considering the largest plausible primary mass in the \gls{LISA} band. For example, taking a maximal source configuration with $M \sim 10^7 M_\odot$ and $\mu \sim 10 M_\odot$ yields $q \sim 10^{-6}$, suggesting that maintaining total flux errors below $\delta \sim 10^{-6}$, ensures waveform suitability across the EMRI parameter space accessible to \gls{LISA}.

These results have immediate implications for the development of higher-order post-adiabatic (PA) models~\cite{Wardell:2021fyy, Burke:2023lno, Mathews:2025nyb}, where higher fidelity in flux modeling and interpolation becomes even more critical. Our findings highlight that systematic waveform errors—originating not from physical approximations but from computational choices—can propagate into biases that are comparable to those introduced by neglected physics. Future EMRI waveform models must therefore address both physical and numerical accuracy to fully realize the science potential of \gls{LISA}.

% \vspace{3.2cm}
% \clearpage
\section{Acknowledgements}
H. K. acknowledges Alessandra Buonanno and the ACR group at the Max Planck Institute for Gravitational Physics (Albert Einstein Institute) for supporting his visit and providing an engaging research environment. H. K. also acknowledges Enrico Barausse and the Astroparticle Physics group at SISSA for supporting his visit and thanks Enrico for insightful discussions on convergence of flux mode contributions. \\
H. K. also gratefully acknowledges the Braithwaite Travel Grant from the University of Guelph, which supported his participation in the \gls{LISA} Symposium (July 2024, Dublin) where preliminary results of this work were presented.\\
Research at Perimeter Institute is supported in part by the Government of Canada through the Department of Innovation, Science and Economic Development and by the Province of Ontario through the Ministry of Colleges and Universities.\\
Additionally, we performed our numerical simulations on the "Symmetry" HPC at the Perimeter Institute.\\
L. S. acknowledges the Perimeter Institute for Theoretical Physics for supporting his visit. L. S. would like to acknowledge the support of the European Space Agency through ESA’s postdoctoral Research Fellowship programme.\\
OB acknowledges financial support from the Grant UKRI972 awarded via the UK Space Agency and computational resources from the French space agency CNES in the framework of \gls{LISA}. \\
MvdM acknowledges financial support by 
the VILLUM Foundation (grant no. VIL37766),
the DNRF Chair program (grant no. DNRF162) by the Danish National Research Foundation and the MPI for Gravitational Physics,
and the European Union’s Horizon ERC Synergy Grant “Making Sense of the Unexpected in the Gravitational-Wave Sky” grant agreement no.\ GWSky–101167314.
\\
ZN acknowledges support from the ERC Consolidator/UKRI Frontier Research Grant GWModels (selected by the ERC and funded by UKRI [grant number EP/Y008251/1]).

\clearpage
\bibliographystyle{IEEEtran}% apsrev4-1
\bibliography{accuracy}

\end{document}

%% file: abbr.tex
\newacronym{GW}{GW}{Gravitational Waves}
\newacronym{NS}{NS}{Neutron Star}
\newacronym{BH}{BH}{Black Hole}
\newacronym{BBH}{BBH}{Binary Black Hole}
\newacronym{BNS}{BNS}{Binary Neutron Star}
\newacronym{BHNS}{BHNS}{Black Hole Neutron Star}
\newacronym{EMRIs}{EMRIs}{Extreme Mass-Ratio Inspirals}
\newacronym{FEW}{FEW}{FastEMRIWaveforms}
\newacronym{PE}{PE}{parameter estimation}
\newacronym{EHT}{EHT}{Event Horizon Telescope}
\newacronym{MCMC}{MCMC}{Markov-Chain Monte-Carlo}
\newacronym{EoS}{EoS}{Equation of State}
\newacronym{GR}{GR}{General Relativity}
\newacronym{PN}{PN}{Post-Newtonian}
\newacronym{EdGB}{EdGB}{Einstein-dilation Gauss-Bonnet}
\newacronym{sGB}{sGB}{Scalar Gauss-Bonnet gravity}
\newacronym{LMXB}{LMXB}{Low-Mass X-ray Binary}
\newacronym{SMBH}{SMBH}{Supermassive Black Holes}
\newacronym{sBH}{sBH}{Stellar-mass Black Hole}
\newacronym{MGO}{MGO}{Mass-Gap Object}
\newacronym{PBH}{PBH}{Primordial Black Holes}
\newacronym{EOB}{EOB}{Effective-One-Body}
\newacronym{PSD}{PSD}{Power Spectral Density}
\newacronym{SPA}{SPA}{Stationary Phase Approximation}
\newacronym{CDF}{CDF}{Cumulative Distribution Function}
\newacronym{SDDR}{SDDR}{Savage-Dickey Density Ratio}
\newacronym{SNR}{SNR}{Signal-to-Noise Ratio}
\newacronym{LIGO}{LIGO}{Laser Interferometer Gravitational-wave Observatory}
\newacronym{Virgo}{Virgo}{Virgo is an interferometric gravitational-wave antenna}
\newacronym{KAGRA}{KAGRA}{KAmioka GRAvitational wave detector}
\newacronym{LISA}{LISA}{Laser Interferometer Space Antennae}
\newacronym{TaiJi}{TaiJi}{a Chinese space-based gravitational wave observatory}
\newacronym{TianQin}{TianQin}{a Chinese space-based gravitational wave observatory}
\newacronym{PyCBC}{PyCBC}{a Python-based package used to detect Compact Binaries Coalescence and measure the astrophysical parameters of detected sources}
\newacronym{TaylorF2}{TaylorF2}{Taylor F2 approximant frequency domain waveform model}
\newacronym{IMRPhenomXPHM}{IMRPhenomXPHM}{a frequency domain Phenomenological waveform model for Processing BBHs at Inspiral-Merger-Ringdown stage}
\newacronym{FIM}{FIM}{Fisher information matrix}
\newacronym{ppE}{ppE}{parameterized post-Einsteinian}
\newacronym{BP}{BP}{Black-Hole Perturbation}
\newacronym{AGN}{AGN}{Active Galactic Nuclei}
\newacronym{ISCO}{ISCO}{Innermost Stable Circular Orbit}
\newacronym{PTAs}{PTAs}{Pulsar Timing Arrays}
\newacronym{CMB}{CMB}{Cosmic Microwave Background}
\newacronym{EM}{EM}{ElectroMagneti}
\newacronym{SGWB}{SGWB}{Stochastic Gravitational-wave Background}
\newacronym{PA}{PA}{post-adiabatic}